\documentclass[a4paper,11pt]{article}
\usepackage{jheppub} % for details on the use of the package, please see the JINST-author-manual
\usepackage[normalem]{ulem}
\usepackage{xcolor, graphicx,import,mathrsfs,braket,tikz,amsmath}
\usepackage[export]{adjustbox}
\usepackage{graphicx}
\usepackage{booktabs}
\usepackage{tabularx}
\usepackage{array}
\usepackage{colortbl}
\usepackage{microtype}
\usetikzlibrary{arrows.meta}
\usetikzlibrary{positioning}

\newcommand{\DefectConstructionDiagram}{%
\begin{tikzpicture}[
    baseline=(diagrambase),
    x=1.1cm,
    y=1.1cm,
    bulk/.style={
        font=\scriptsize,
        align=center
    },
    defect/.style={
        font=\scriptsize,
        align=center
    },
    annotation/.style={
        font=\footnotesize,
        align=center
    }
]
    \coordinate (diagrambase) at (1.9,.57);

    \path[
        fill=black!6,
        draw=black!14,
        line width=.35pt,
        rounded corners=.7pt
    ]
        (0,0) rectangle (4,1.14);

    \draw[purple!72,line width=1pt]
        (2,.01) -- (2,1.13);

    \node[bulk] at (1,.57)
        {$X_L$\\[-1pt]$(g_L,H_L)$};

    \node[bulk] at (3,.57)
        {$X_R$\\[-1pt]$(g_R,H_R)$};

    \node[defect,anchor=south] at (2,1.18)
        {$(Y,F)$};
\end{tikzpicture}%
}

\preprint{{IFT-UAM/CSIC-26-108}\\}

\newcommand{\RawGeometricFusionDiagram}{%
\begin{tikzpicture}[
    baseline=(fusionbase),
    x=1.cm,
    y=1.cm,
    bulk/.style={
        font=\scriptsize,
        align=center
    },
    defect/.style={
        font=\scriptsize,
        align=center
    },
    constraint/.style={
        font=\footnotesize
    }
]
    \coordinate (fusionbase) at (3,.57);

    \path[
        fill=black!6,
        draw=black!14,
        line width=.35pt,
        rounded corners=.7pt
    ]
        (0,0) rectangle (5.8,1.14);

    \draw[orange!82,line width=.95pt]
        (1.93,.01) -- (1.93,1.13);
    \draw[blue!72,line width=.95pt]
        (3.9,.01) -- (3.9,1.13);

    \node[bulk] at (1,.57) {$X_L$};
    \node[bulk] at (3,.57) {$X_I$};
    \node[bulk] at (5,.57) {$X_R$};

    \node[defect,anchor=south] at (1.6,1.18)
        {$(Y_{LI},\,F_{LI},\,\nabla_{LI})$};

    \node[defect,anchor=south] at (4.4,1.18)
        {$(Y_{IR},\,F_{IR},\,\nabla_{IR})$};

    \node[constraint,anchor=north] at (3,-.13)
        {$Y_{LR}^{\rm raw}
          =Y_{LI}\times_{X_I}Y_{IR}$};
\end{tikzpicture}%
}

\newcommand{\GeometricFusionArrow}{%
\begin{tikzpicture}[
    baseline=(fusionbase),
    x=1.1cm,
    y=1.1cm
]
    \coordinate (fusionbase) at (1.55,.57);

    \draw[
        ->,
        line width=.75pt,
        draw=black!70
    ]
        (0,.57) -- (1.6,.57);

    \node[
        font=\scriptsize,
        anchor=south
    ] at (0.8,.66)
        {fusion};    

\end{tikzpicture}%
}

\newcommand{\FusedGeometricFusionDiagram}{%
\begin{tikzpicture}[
    baseline=(fusionbase),
    x=1.1cm,
    y=1.1cm,
    bulk/.style={
        font=\scriptsize,
        align=center
    },
    defect/.style={
        font=\scriptsize,
        align=center
    },
    constraint/.style={
        font=\footnotesize
    }
]
    \coordinate (fusionbase) at (2,.57);

    \path[
        fill=black!6,
        draw=black!14,
        line width=.35pt,
        rounded corners=.7pt
    ]
        (0,0) rectangle (3,1.14);

    \draw[purple!72,line width=1pt]
        (1.5,.01) -- (1.5,1.13);

    \node[bulk] at (0.7,.57) {$X_L$};
    \node[bulk] at (2.3,.57) {$X_R$};

    \node[defect,anchor=south] at (1.5,1.18)
        {$(Y_{LR}^{(a)},\,F_{LR}^{(a)},\,\nabla_{LR}^{(a)})$};

    \node[constraint,anchor=north] at (1.5,-.1)
        {$\pi_a^*F_{LR}^{(a)}
          =F_{LR}^{\rm raw}|_{Z_a}$};
\end{tikzpicture}%
}

\newcommand{\RawFusionDiagram}{%
\begin{tikzpicture}[
    baseline=(fusionbase),
    x=.9cm,
    y=.9cm,
    bulk/.style={
        font=\small
    },
    defect/.style={
        font=\scriptsize
    },
    constraint/.style={
        font=\footnotesize
    }
]
    \coordinate (fusionbase) at (3,.57);

    \path[
        fill=black!6,
        draw=black!14,
        line width=.35pt,
        rounded corners=.7pt
    ]
        (0,0) rectangle (6,1.14);

    \draw[orange!82,line width=.95pt]
        (2,.01) -- (2,1.13);
    \draw[blue!72,line width=.95pt]
        (4,.01) -- (4,1.13);

    \node[bulk] at (1,.57) {$\omega_1$};
    \node[bulk] at (3,.57) {$\omega_2$};
    \node[bulk] at (5,.57) {$\omega_3$};

    \node[defect,anchor=south] at (2,1.18) {$\alpha$};
    \node[defect,anchor=south] at (4,1.18) {$\beta$};

    \node[constraint,anchor=north] at (3,-.13)
        {$q_{12}=g_1g_2^{-1}
          \qquad
          q_{23}=g_2g_3^{-1}$};
\end{tikzpicture}%
}

\newcommand{\FusedFusionDiagram}{%
\begin{tikzpicture}[
    baseline=(fusionbase),
    x=.9cm,
    y=.9cm,
    bulk/.style={
        font=\small
    },
    defect/.style={
        font=\scriptsize
    },
    constraint/.style={
        font=\footnotesize
    }
]
    \coordinate (fusionbase) at (2,.57);

    \path[
        fill=black!6,
        draw=black!14,
        line width=.35pt,
        rounded corners=.7pt
    ]
        (0,0) rectangle (4,1.14);

    \draw[purple!72,line width=1pt]
        (2,.01) -- (2,1.13);

    \node[bulk] at (1,.57) {$\omega_1$};
    \node[bulk] at (3,.57) {$\omega_3$};

    \node[defect,anchor=south] at (2,1.18)
        {$\gamma\in I_{\alpha,\beta}$};

    \node[constraint,anchor=north] at (2,-.13)
        {$q_{13}=g_1g_3^{-1}$};
\end{tikzpicture}%
}

\newcommand{\FusionArrow}{%
\begin{tikzpicture}[baseline=-.55ex]
    \draw[
        -{Stealth[length=3.4pt,width=4.2pt]},
        black!65,
        line width=.55pt
    ]
        (0,0) -- (1.55,0)
        node[
            midway,
            above=2.5pt,
            font=\scriptsize,
            text=black
        ]
        {fusion};
\end{tikzpicture}%
}

\newcommand{\RawHopfFusionDiagram}{%
\begin{tikzpicture}[
    baseline=(fusionbase),
    x=1.cm,
    y=1.cm,
    bulk/.style={
        font=\scriptsize,
        align=center
    },
    defect/.style={
        font=\scriptsize
    },
    constraint/.style={
        font=\footnotesize
    }
]
    \coordinate (fusionbase) at (3,.57);

    \path[
        fill=black!6,
        draw=black!14,
        line width=.35pt,
        rounded corners=.7pt
    ]
        (0,0) rectangle (6,1.14);

    \draw[orange!82,line width=.95pt]
        (2,.01) -- (2,1.13);
    \draw[blue!72,line width=.95pt]
        (4,.01) -- (4,1.13);

    \node[bulk] at (1,.57)
        {$X_L$\\[-1pt]$H_L=0$};

    \node[bulk] at (3,.57)
        {$\widetilde X_I$\\[-1pt]$H_{\widetilde I}=p$};

    \node[bulk] at (5,.57)
        {$X_R$\\[-1pt]$H_R=0$};

    \node[defect,anchor=south] at (2,1.18)
        {$T$};

    \node[defect,anchor=south] at (4,1.18)
        {$T^\vee$};

    \node[constraint,anchor=north] at (3,-.13)
        {$\theta_L=\theta_I=\theta_R
          \qquad
          \psi_L=\psi_I=\psi_R$};
\end{tikzpicture}%
}

\newcommand{\FusedHopfFusionDiagram}{%
\begin{tikzpicture}[
    baseline=(fusionbase),
    x=1.cm,
    y=1.cm,
    bulk/.style={
        font=\scriptsize,
        align=center
    },
    defect/.style={
        font=\scriptsize
    },
    constraint/.style={
        font=\footnotesize
    }
]
    \coordinate (fusionbase) at (2,.57);

    \path[
        fill=black!6,
        draw=black!14,
        line width=.35pt,
        rounded corners=.7pt
    ]
        (0,0) rectangle (4,1.14);

    \draw[purple!72,line width=1pt]
        (2,.01) -- (2,1.13);

    \node[bulk] at (1,.57)
        {$X_L$\\[-1pt]$H_L=0$};

    \node[bulk] at (3,.57)
        {$X_R$\\[-1pt]$H_R=0$};

    \node[defect,anchor=south] at (2,1.18)
        {$T\star T^\vee$};

    \node[constraint,anchor=north] at (2,-.13)
        {$\phi_L
          =\phi_R-\dfrac{2\pi}{p}\eta,
          \quad
          \eta\in\mathbb Z_p$};
\end{tikzpicture}%
}

\newcommand{\RawTsTFusionDiagram}{%
\begin{tikzpicture}[
    baseline=(fusionbase),
    x=.82cm,
    y=1cm,
    bulk/.style={
        font=\scriptsize,
        align=center
    },
    defect/.style={
        font=\scriptsize
    }
]
    \coordinate (fusionbase) at (4,.57);

    \path[
        fill=black!6,
        draw=black!14,
        line width=.35pt,
        rounded corners=.7pt
    ]
        (0,0) rectangle (8,1.14);

    \draw[orange!82,line width=.95pt]
        (2,.01) -- (2,1.13);
    \draw[green!55!black,line width=.95pt]
        (4,.01) -- (4,1.13);
    \draw[blue!72,line width=.95pt]
        (6,.01) -- (6,1.13);

    \node[bulk] at (1,.57)
        {$X_L$\\[-1pt]$H_L=0$};

    \node[bulk] at (3,.57)
        {$\widetilde X_I$};

    \node[bulk] at (5,.57)
        {$\widetilde X_M$};

    \node[bulk] at (7,.57)
        {$X_{\zeta,R}$\\[-1pt]$H_\zeta$};

    \node[defect,anchor=south] at (2,1.18)
        {$T$};

    \node[defect,anchor=south] at (4,1.18)
        {$S_\zeta$};

    \node[defect,anchor=south] at (6,1.18)
        {$T_\zeta^\vee$};
\end{tikzpicture}%
}

\newcommand{\FusedTsTFusionDiagram}{%
\begin{tikzpicture}[
    baseline=(fusionbase),
    x=.82cm,
    y=1cm,
    bulk/.style={
        font=\scriptsize,
        align=center
    },
    defect/.style={
        font=\scriptsize
    }
]
    \coordinate (fusionbase) at (2,.57);

    \path[
        fill=black!6,
        draw=black!14,
        line width=.35pt,
        rounded corners=.7pt
    ]
        (0,0) rectangle (4,1.14);

    \draw[purple!72,line width=1pt]
        (2,.01) -- (2,1.13);

    \node[bulk] at (1,.57)
        {$X_L$\\[-1pt]$H_L=0$};

    \node[bulk] at (3,.57)
        {$X_{\zeta,R}$\\[-1pt]$H_\zeta$};

    \node[defect,anchor=south] at (2,1.18)
        {$Y_\zeta^{\mathrm{TsT}}$};
\end{tikzpicture}%
}

\newcommand{\RawTdualBiconjFusionDiagram}{%
\begin{tikzpicture}[
    baseline=(fusionbase),
    x=1.1cm,
    y=1.1cm,
    bulk/.style={
        font=\scriptsize,
        align=center
    },
    defect/.style={
        font=\scriptsize
    },
    constraint/.style={
        font=\footnotesize
    }
]
    \coordinate (fusionbase) at (3,.57);

    \path[
        fill=black!6,
        draw=black!14,
        line width=.35pt,
        rounded corners=.7pt
    ]
        (0,0) rectangle (6,1.14);

    \draw[blue!72,line width=.95pt]
        (2,.01) -- (2,1.13);

    \draw[orange!82,line width=.95pt]
        (4,.01) -- (4,1.13);

    \node[bulk] at (1,.57)
        {$X_L$\\[3pt]$H_L=H$};

    \node[bulk] at (3,.57)
        {$X_I$\\[3pt]$H_I=H$};

    \node[bulk] at (5,.57)
        {$\widetilde X_R$\\[3pt]$H_{\widetilde R}=\widetilde H$};

    \node[defect,anchor=south] at (2,1.18)
        {$B_\alpha$};

    \node[defect,anchor=south] at (4,1.18)
        {$T$};

    \node[constraint,anchor=north] at (3,-.13)
        {$g_Lg_I^{-1}\in C_\alpha
          \qquad
          \pi(g_I)=\pi(\widetilde x_R)$};
\end{tikzpicture}%
}

\newcommand{\FusedTdualBiconjFusionDiagram}{%
\begin{tikzpicture}[
    baseline=(fusionbase),
    x=1.1cm,
    y=1.1cm,
    bulk/.style={
        font=\scriptsize,
        align=center
    },
    defect/.style={
        font=\scriptsize
    },
    constraint/.style={
        font=\footnotesize
    }
]
    \coordinate (fusionbase) at (2,.57);

    \path[
        fill=black!6,
        draw=black!14,
        line width=.35pt,
        rounded corners=.7pt
    ]
        (0,0) rectangle (4,1.14);

    \draw[purple!72,line width=1pt]
        (2,.01) -- (2,1.13);

    \node[bulk] at (1,.57)
        {$X_L$\\[3pt]$H_L=H$};

    \node[bulk] at (3,.57)
        {$\widetilde X_R$\\[-1pt]$H_{\widetilde R}=\widetilde H$};

    \node[defect,anchor=south] at (2,1.18)
        {$Y_{\alpha T}^{\pm}$};

    \node[constraint,anchor=north] at (2,-.13)
        {$U_{\alpha T}:\ |\cos\alpha|<R_{LR}$};
\end{tikzpicture}%
}

\newcommand{\TdualBiconjSetupDiagram}{%
\begin{tikzpicture}[baseline=(setupbase),x=1.3cm,y=1.1cm]
    \coordinate (setupbase) at (3,.7);

    \path[fill=black!6,rounded corners=2pt]
        (0,0) rectangle (6,1.4);

    \draw[orange!82,line width=.95pt] (2,0) -- (2,1.4);
    \draw[blue!72,line width=.95pt] (4,0) -- (4,1.4);

    \node[font=\footnotesize,align=center] at (1,.7)
        {$X_L$\\[5pt]$SU(2)_k\ \mathrm{WZW}$};

    \node[font=\footnotesize,align=center] at (3,.7)
        {$X_I$\\[5pt]$SU(2)_k\ \mathrm{WZW}$};

    \node[font=\footnotesize,align=center] at (5,.7)
        {$\widetilde X_R$\\[3.5pt]$L(k,1)$};

    \node[font=\footnotesize,above=3pt] at (2,1.35)
        {$(B_\alpha, F_\alpha)$};

    \node[font=\footnotesize,above=3pt] at (4,1.35)
        {$(Y_T,F_T)$};
\end{tikzpicture}%
}

\title{\boldmath Target-space fusion of classical topological defects}

\author[a]{Saskia Demulder,}
\affiliation[a]{Department of Quantitative Methods, CUNEF Universidad,\\ Calle Almansa 101, 28040 Madrid, Spain}
\emailAdd{saskia.demulder@cunef.edu}

\author[b,c]{Chuying Wang\,}
\affiliation[b]{Instituto de F\'isica Te\'orica IFT-UAM/CSIC,\\ Calle Nicol\'as Cabrera 13-15, Campus de Cantoblanco, 28049 Madrid, Spain}

\affiliation[c]{Departamento de F\'{i}sica Te\'{o}rica, Universidad Aut\'{o}noma de Madrid, \\Cantoblanco, 28049 Madrid, Spain} 
\emailAdd{chuying.wang@ift.csic.es}

\abstract{Classical fusion of sigma-model defects is commonly approached by combining their worldsheet actions and eliminating the intermediate fields. This procedure, however, does not systematically identify the resulting target-space defect or its possible branches. We address this problem by formulating fusion as a reduction of the target-space defect data. Local descent of the intrinsic two-form and global descent of the connection determine the admissible branches. For symmetry-preserving bi-branes in the $SU(2)$ WZW model and Hopf-fibre T-duality defects, the prescription reproduces the known biconjugacy-class branches and the finite family of fibre-shift defects. The appearance of several branches provides a classical manifestation of non-invertible fusion. We further use the framework to construct TsT and T-dual biconjugacy defects, including their global consistency conditions. It therefore provides a systematic method for computing classical fusion and constructing new defects from known ones.}

\begin{document}
 \maketitle
\flushbottom

\newpage
%%%%%%%%%%%%%%%%%%%%%%%%%%%%%%%%%%%%%%%%%%%%%%%%%%%%%%
%%%%%%%%%%%%%%         Introduction       %%%%%%%%%%%%
%%%%%%%%%%%%%%%%%%%%%%%%%%%%%%%%%%%%%%%%%%%%%%%%%%%%%%
\section{Introduction}

Generalised symmetries have substantially broadened the role of symmetry in quantum field theory and quantum many-body systems
\cite{Gaiotto:2014kfa,McGreevy:2022oyu,Shao:2023gho,Bhardwaj:2023kri,schafernameki2023ictplecturesnoninvertiblegeneralized,Luo_2024}. They are implemented by topological operators of different codimensions. In two dimensions these operators are topological lines, that is of co-dimension one. When placed along a spatial slice, they define maps between Hilbert spaces, while bringing two such lines together defines their fusion. Such lines need not be invertible and their fusion may decompose into several distinct defects. Their fusion (and additional information such as junctions) is consequently organised by tensor categories rather than groups \cite{Petkova:2000ip,Frohlich:2003hm,Frohlich:2004ef,Fuchs:2007tx}. This categorical description has become a central tool for studying generalised and non-invertible symmetries. There is also a classical counterpart in terms of topological defect actions and their gluing conditions, but its fusion structure is considerably less systematic.

In this paper we precisely study the classical counterpart of fusion and realisation of line defects in two-dimensional non-linear sigma models. A classical defect\footnote{The term ``interface'' is often reserved for a defect separating two different theories, while ``defect'' may refer specifically to the case in which the theories on both sides coincide. We use ``defect'' for both situations throughout.} separates worldsheet regions with target-spaces $X_L$ and $X_R$ and contributes a term supported on the defect line. 
\begin{equation}\label{eq:defect-action-intro}
S_{\mathrm{tot}}=S_L[\phi_L]+S_R[\phi_R]+\int_D \Phi_D^*A\,.
\end{equation}
Here $D$ is the oriented defect line, while
$\Phi_D=(\phi_L,\phi_R)|_D$ is the restriction of the two sigma-model fields to $D$ and takes values in the defect worldvolume inside $X_L\times X_R$. The one-form $A$ specifies the local coupling carried by the defect.
When varying the complete action, one determines the gluing conditions relating the fields on its two sides. The defect is conformal when the energy flux is continuous across the line and topological when the two chiral components of the stress tensor are separately continuous. In the latter case, when there is no obstruction, the defect can be deformed without changing classical observables. From the Hamiltonian perspective, one can see such defects as defining Lagrangian correspondences between the phase spaces of the two theories \cite{Arvanitakis:2023sho,Arvanitakis:2024vhz}; for a closely related formulations in terms of Dirac structures see \cite{Demulder:2022nlz}.

A setting in which classical topological defects have been studied comparatively extensively is T-duality. Recent constructions derive the corresponding defects directly from their worldsheet actions \cite{Niro:2022ctq}, including through half-space gauging \cite{Arias-Tamargo:2025xdd,Arias-Tamargo:2025fhv}, while Poisson-Lie T-duality has been formulated using both worldsheet and target-space data \cite{Demulder:2022nlz}.  The defect description of T-duality is captured locally by the Poincar\'e line bundle \cite{Bouwknegt:2003vb} and has been studied for toroidal sigma models, compact bosons and WZW backgrounds \cite{Fuchs:2007tx,Sarkissian:2008dq,Kapustin:2010zc,
Elitzur:2013ut}. At the quantum level, topological T-duality defects were constructed for the compact boson and for $SU(2)$ lens-space theories in \cite{Fuchs:2007tx,Sarkissian:2008dq}.

To formulate and organise their classical fusion, we adopt the target-space perspective and build on the bi-brane geometry and topologicality conditions developed in \cite{Fuchs:2007fw,Kapustin:2010zc}.\footnote{Other systematic constructions include defects in rational conformal field theory described by bimodules, phase-transition defects in gauged linear sigma models \cite{Brunner:2021ulc,Brunner:2021cga}, and in SCFTs and their topological B-twist \cite{Caldararu:2025eoj}. Related dualities and their defects also admit lattice realisations in compact-boson and Villain-type models \cite{Gorantla:2021svj,Cheng:2022sgb,Pace:2024oys, Argurio:2026txf,Bacq:2026ega}.}
Sigma models with Wess-Zumino terms also admit a target-space description of its defects. The values of the fields along the defect define a submanifold
\begin{equation}
Y\subseteq X_L\times X_R\,,
\end{equation}
equipped with an intrinsic two-form $F$ satisfying
\begin{equation}
dF=p_L^*H_L-p_R^*H_R\,.
\end{equation}
On local bulk patches, with $db_L=H_L$ and $db_R=H_R$, the one-form $A$ entering the defect action is related to $F$ by
\begin{equation}
F=p_L^*b_L-p_R^*b_R+dA\,.
\end{equation}
Thus $Y$ specifies the allowed values of the fields along the defect, while $F$ combines the local defect coupling with its compatibility with the two bulk Wess--Zumino terms. These data were introduced as a bi-brane in \cite{Fuchs:2007fw}. In \cite{Kapustin:2010zc}, the authors subsequently translated the conformal and topological gluing conditions into geometric conditions on $Y$, the target-space metrics and $F$. This makes it then possible to construct and test classical defects directly from their target-space data, rather than starting from a proposed defect action.

Fusion at the classical level is deceptively simple. Consider two defects placed at distinct spatial slices, separating theories with target spaces $X_L$, $X_I$ and $X_R$. Locally, their worldsheet action takes the form
\begin{equation}\label{eq:two-defect-action-intro}
S_{\mathrm{tot}}=S_L[\phi_L]+S_I[\phi_I]+S_R[\phi_R]+\int_{D_{LI}}\Phi_{D_{LI}}^*A_{LI}+\int_{D_{IR}}\Phi_{D_{IR}}^*A_{IR}\,.
\end{equation}
For topological defects the displacement operator vanishes, so the two defect lines may be brought together without changing classical observables. At the level of the action, fusion then appears to amount to shrinking the intermediate region and combining the two defect couplings. This can be carried out directly in particular examples, as for the classical fusion of T-duality defects \cite{Niro:2022ctq}, whose result agrees with the corresponding quantum fusion rule \cite{Fuchs:2007tx,Thorngren:2021yso}. For generic conformal interfaces, by contrast, fusion may require a short-distance regularisation and involve a defect renormalisation-group flow \cite{Bachas:2007td}.

Classical fusion is nevertheless not determined simply by adding the two worldsheet actions. After two defect lines are brought together, the resulting action still contains the field in the intermediate region and does not by itself identify the defect, or collection of defects, obtained after that field is eliminated. Target-space composition by fibre product and projection was already discussed for bi-branes in \cite{Fuchs:2007fw}.
Related global aspects of WZW defects, including defect networks and the geometric description of affine $\widehat{\mathfrak{su}}(2)_k$ fusion, were studied in \cite{Runkel:2008gr,Runkel:2009sp}. Our present aim is to exploit this target-space picture in order to formulate an explicit classical fusion prescription. That is, to determine which defect branches arise after eliminating the intermediate field and which two-form and connection data they carry.

Our starting point is that the target-space geometry should guide the elimination of the intermediate field. We first compose the defect supports over the common target and combine their differential data. We then determine where these data are independent of the intermediate field and descend to the projected support. This descent has both a local part, expressed through the intrinsic two-form, and a global part, detected by the holonomy of the composite connection. The admissible loci obtained in this way determine the resulting fusion branches. Thus the support, defect coupling and possible decomposition into several defects are obtained within a single geometric procedure.

We illustrate the prescription through examples that both recover known fusion rules and use fusion to construct new classical defects. We first apply the prescription to the symmetry-preserving $SU(2)$ WZW bi-branes. This reproduces the known biconjugacy-class composition \cite{Fuchs:2007fw} (see also \cite{Klimcik:2013rla}), while making the descent of the two-form and connection data explicit. We then consider the Hopf-fibre T-duality defect and its reverse. Their composition produces the finite family of fibre-shift defects derived in \cite{Niro:2022ctq,Bharadwaj:2024gpj}, with the $\mathbb Z_p$ branches selected geometrically by fibre holonomy. We further construct defects implementing TsT transformations and compose a symmetry-preserving WZW bi-brane with a WZW-compatible Hopf defect. Note that quantum T-dual WZW defects were constructed in \cite{Sarkissian:2008dq}. Together, these examples yield the classical composition rules summarised in Table \ref{tab:classical-defect-fusion}.

One advantage of taking a geometric approach to classical fusion is that, besides formalising fusion beyond the naive ``summation'' of defect actions, it opens a possible route towards quantisation and towards relating the classical and quantum descriptions of topological defects. Although we hope that our results clarify some of the geometric input required for this problem, quantisation lies beyond the scope of this paper. For WZW bi-branes, the relation to quasi-Hamiltonian quantisation and Chern-Simons theory is well developed; see, for example, \cite{Axelrod:1989xt,alekseev1998lie,Sarkissian:2009hy, meinrenken2006lectures,meinrenken2011quantization}. Already for T-duality defects, however, the corresponding problem is more subtle because the quantum construction involves orbifold and twisted sectors. We return briefly to this distinction in the outlook.

The paper is organised as follows. Section \ref{sec:review} reviews the target-space description of sigma-model defects and the geometric conformality and topologicality conditions. In Section \ref{sec:classical_fusion} we formulate the connection-level fusion prescription. Section \ref{sec:constructions} applies it to symmetry-preserving WZW bi-branes, Hopf-fibre T-duality, shift and TsT defects, and T-dual biconjugacy correspondences. Section \ref{sec:conclusions} summarises the resulting composition laws and discusses open directions. Technical topologicality checks and further ordered fusion products are collected in the appendices.

%%%%%%%%%%%%%%%%%%%%%%%%%%%%%%%%%%%%%%%%%%%%%%%%%%%%%%
%%%%%%%%%%%%%%         Review        %%%%%%%%%%%%%%%%%
%%%%%%%%%%%%%%%%%%%%%%%%%%%%%%%%%%%%%%%%%%%%%%%%%%%%%%
\section{Review: Target-space geometry of classical defects}
\label{sec:review}

This section reviews the target-space formulation of codimension-one defects in two-dimensional sigma models and fixes the conventions used below. Starting from the worldsheet action, including the Wess-Zumino term and the coupling supported on the defect line, we explain how the defect is encoded geometrically by a submanifold  together with a two-form. This target-space description was introduced in the language of bi-branes in \cite{Fuchs:2007fw} and further developed in \cite{Kapustin:2010zc}.\footnote{Terminology varies in the literature. In this paper we reserve ``bi-brane'' for the symmetry-preserving WZW defects considered in Section \ref{sec:constructions}; for general target-space data $(Y,F)$ we use ``defect''.} We also review how the two-form $F$, which form the second half of the target-space information capturing the defect, to the local one-form entering the defect action. This is a distinction that is essential in the presence of non-trivial Wess-Zumino flux and, in particular, for the WZW examples considered in Section \ref{sec:constructions}.

We then review how the usual worldsheet conditions on the energy-momentum tensor translate into conditions on the target-space data $(Y,F)$. Continuity of the energy flux gives the conformal defect condition, while separate continuity of the two chiral components gives the topological condition. Following \cite{Kapustin:2010zc}, we present these conditions first in terms of the defect gluing relations and then in their intrinsic geometric form. This will be the framework and associated criteria that we will be using throughout the paper to construct and use to verify classical conformal and topological defects.

%%%%%%%%%%%%%%%%%%%%%%%%%%%%%%%%%%%%%%%%%%%%%%%%%%%%%%
%%%%%%%%%%%%%%%%%%%%%%%%%%%%%%%%%%%%%%%%%%%%%%%%%%%%%%
\subsection{WZW models and codimension-one defects}\label{sec:WZW-models-defects}

Let $G$ be compact, connected, simple and simply connected, equipped with a fixed invariant bilinear form normalised so that the Cartan three-form represents an integral cohomology class. In the rest of the note we will mostly take $G=SU(2)$.
The WZW model is a two-dimensional non-linear sigma model whose
field content is a map $g:\Sigma_{\mathrm{ws}}\to G$, where $\Sigma_{\mathrm{ws}}$ is the worldsheet and $G$ is a compact Lie group. The action is the sum of a sigma-model kinetic term and a Wess-Zumino term\footnote{Let $G$ be compact, connected, simple and simply connected. In the applications below we take $G=SU(2)$. For $SU(2)$, let $\mathrm{tr}$ denote the ordinary trace in the fundamental representation and choose the orientation so that
\begin{equation}\label{eq:WZW-integral-normalisation}
\frac{1}{24\pi^2} \int_{SU(2)} \mathrm{tr}\bigl(g^{-1}dg\bigr)^3=1\,.
\end{equation}
We use the invariant bilinear form
\begin{equation}\label{eq:inv_pairing_conv}
\langle X,Y\rangle = \frac{1}{2\pi}\mathrm{tr}(XY)\,.
\end{equation}
It is negative definite on $\mathfrak{su}(2)$; the corresponding Riemannian group metric is proportional to $-\langle\,\cdot ,\,\cdot\rangle$.},
\begin{equation}\label{eq:wzw-action}
S[g]=-\frac{k}{8\pi}\int_{\Sigma_{\mathrm{ws}}}\mathrm{tr}\left(g^{-1}dg\wedge\star g^{-1}dg\right)+\frac{k}{12\pi}\int_{\mathcal B}\mathrm{tr}\left(g^{-1}dg\right)^3\,.
\end{equation}
where $\mathcal B$ is a three-manifold with $\partial \mathcal  B=\Sigma_{\mathrm{ws}}$. The closed three-form $H=\frac{k}{12\pi}\mathrm{tr} \left(g^{-1}\mathrm  dg\right)^3$ is the Wess-Zumino flux. The exponentiated Wess-Zumino term is independent of the choice of $\mathcal B$ provided $k\in\mathbb Z$. Classically, the model is invariant under independent left and right multiplication,
\begin{equation}\label{eq:global-LR}
g(x)\mapsto h_L\, g(x)\, h_R^{-1}\,, \qquad h_L,h_R\in G\,.
\end{equation}
Oftentimes, for generic sigma-models, we denote target-space maps by $\phi$. When the target is a Lie group, we use the conventional group-valued notation $g$ for the same field; accordingly, the fields on the two sides of a WZW defect will be denoted by $g_L$ and $g_R$. Although the geometric discussion extends to compact simple groups, all explicit formulas in this example and constructions derived below use $G=SU(2)$.

We are interested in codimension-one defects in such theories. Let $D\subset\Sigma_{\mathrm{ws}}$ be an oriented defect line with $D=\partial\Sigma_L=-\partial\Sigma_R$. The fields on the two sides take values in targets $X_L$ and $X_R$, and their restrictions to the defect define the map
\begin{equation}\label{eq:target-space-map}
\Phi_D := \left.(\phi_L,\phi_R)\right|_D : D\longrightarrow Y\,, \qquad Y\subseteq X_L\times X_R\,.
\end{equation}
Locally, after choosing representatives for the bulk Wess-Zumino data, the action contains the defect coupling
\begin{equation}\label{eq:defect-action}
S_{\mathrm{tot}} = S_L[\phi_L]+S_R[\phi_R] +\int_D\Phi_D^*A\,.
\end{equation}
Here $A$ is a local one-form on $Y$. On local bulk patches choose two-form potentials $db_L=H_L$ and $db_R=H_R$. The gauge-invariant intrinsic two-form on the defect worldvolume is\footnote{In \cite{Fuchs:2007fw} the notation $\varpi$ is used, here instead we denote this two-form by $F$ throughout.}
\begin{equation}\label{eq:intrinsic-defect-form}
F=p_L^*b_L-p_R^*b_R+dA\,,
\end{equation}
and consequently
\begin{equation}\label{eq:bibrane-curvature}
dF=p_L^*H_L-p_R^*H_R\,.
\end{equation}
Under
\begin{equation}
b_i\longmapsto b_i+d\Lambda_i\,, \quad A\longmapsto A-p_L^*\Lambda_L+p_R^*\Lambda_R+d\chi\,,
\end{equation}
the intrinsic form $F$ is invariant.

Thus, the defect action term \eqref{eq:defect-action}, becomes in terms of the local target-space data an embedded subspace together with a two-form defined on it:
\begin{equation}\label{eq:local-defect-data}
Y\hookrightarrow X_L\times X_R\,, \quad F\in\Omega^2(Y)\,,
\end{equation}
where $F$ satisfies \eqref{eq:bibrane-curvature}. This will be the main language in which we will consider fusion and construction of defects in what follows.

Equations \eqref{eq:intrinsic-defect-form} and \eqref{eq:bibrane-curvature} express the compatibility of the defect data with the bulk Wess-Zumino terms, but do not by themselves imply topologicality. In fact up to now we did not impose or require any specific gluing conditions. Within the class of defect actions \eqref{eq:defect-action}, the gluing relation obtained by varying the action automatically implies continuity of the classical energy flux and hence conformality. Topologicality imposes the additional conditions reviewed in the next subsection. Before discussing those, let us comment a little further on what this geometric data is.

Globally, the local one-form $A$ in the defect action has now been effectively replaced by compatible rank-one\footnote{By rank-one connection data we mean a single $U(1)$ connection, locally represented by the one-form $A$. If several sheets of a covering are retained without being identified, one instead obtains one copy of these data from each sheet. This distinction will become important when considering concrete examples.} connection data associated, which we denote by $\nabla$. In the presence of Wess-Zumino flux these data is twisted but the associated flux, and $F$ need not be the curvature of an ordinary line bundle. 
Since we dispose of a manifold\footnote{Generically $Y$ need not, and in fact will not be as a result fusion in certain cases, be a manifold. See further comments below.} together with a (twisted) connection, one can call the data $(Y,F)$ prequantisable when it admits such a global refinement with the required integrality properties. Note that this is not very different from quantisable D-branes, as for example considered in \cite{Alekseev:1998mc}. Hence, in what follow, we will sometimes use this terminology to mean that the geometric input relevant for a potential future quantisation; though, as mentioned earlier, we will not attack the challenge of quantising defect in the sigma models here.

%%%%%%%%%%%%%%%%%%%%%%%%%%%%%%%%%%%%%%%%%%%%%%%%%%%%%%
%%%%%%%%%%%%%%%%%%%%%%%%%%%%%%%%%%%%%%%%%%%%%%%%%%%%%%
\subsection{Geometric conformal and topological conditions}\label{subsec:defect conf-topo-conditions}

We now review the geometric characterisation of bosonic conformal and topological defects conditions as derived in \cite{Kapustin:2010zc}. The starting point is the usual worldsheet definition in terms of continuity of the stress-energy tensor across the defect line. That is the imposition of particular continuity relation between the energy momentum tensors of the theories that are being divided by the defect. How the energy flows across the defect, directly determines its property.  In Lorentzian worldsheet coordinates, 
\begin{align}\label{eq:conformal-topological-conditions}
\text{conformal}:&\qquad  T^1{}_0-\widehat T^1{}_0=0\,,\\
\text{topological}:&\qquad 
T^1{}_0-\widehat T^1{}_0=0\,, \qquad T^1{}_1-\widehat T^1{}_1=0\,,
\end{align}
where the stress-energy tensor of a sigma model is given by
\begin{equation}
    T_{\mu\nu}= g_{ij}(X)\partial_\mu X^i\partial_\nu X^j-\frac{1}{2}g^\mathrm{ws}_{\mu\nu}g^{\gamma\delta}_\mathrm{ws} \, g_{ij}(X)\partial_\gamma X^i\partial_\delta X^j\,,
\end{equation}
where $g^{\mathrm{ws}}_{\mu\nu}$ is the worldsheet metric and $g_{ij}$ is the target-space metric.

%%%%--------- figure ---------%%%%
\begin{figure}[t]
    \centering
    \includegraphics[width=0.7\linewidth]{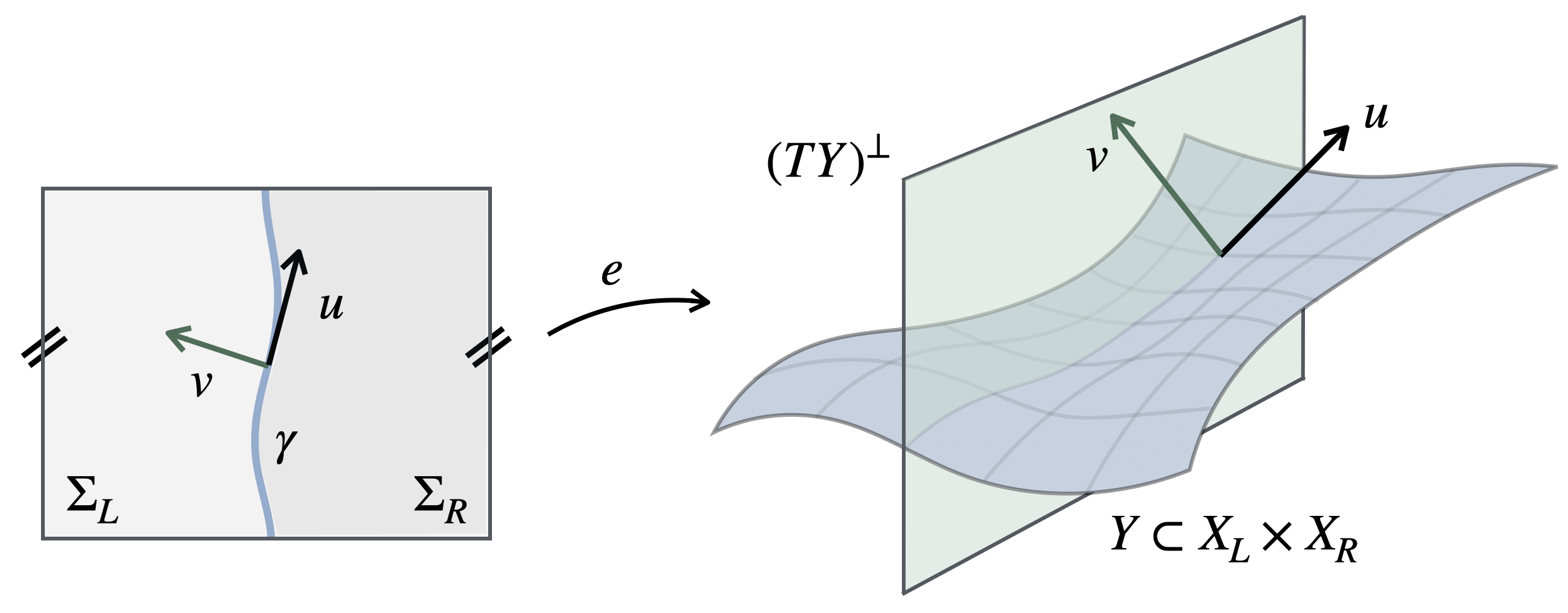}
    \caption{
   The defect map $\Phi_D$ takes the defect line $D$ to the support $Y\subset X_L\times X_R$. Its tangential derivative defines $u\in TY$, while $v$ collects the normal derivatives from the two sides. The bundle $(TY)^\perp$ is the $G$-orthogonal bundle of $TY$.}
    \label{fig:tangent_space_Y}
\end{figure}
%%%%--------- figure ---------%%%%

Now following \cite{Kapustin:2010zc}, we review how these conditions naturally become constrains on the geometric data $(Y,F)$ associated to a sigma-model defect. To express the stress-tensor conditions geometrically, equip the $M:=X_L\times X_R$ product target
\begin{equation}\label{eq:neutral_metric}
G:=g_L\oplus(-g_R).
\end{equation}
If $\dim X_L=\dim X_R=n$, then $G$ has split signature $(n,n)$. This is crucial: unlike the positive-definite case, $G$ can have nontrivial null directions along a submanifold $Y$. For a submanifold $Y\subseteq M$, its $G$-orthogonal bundle is defined by
\begin{equation}
(TY)^\perp :=\bigl\{\xi\in T(X_L\times X_R)|_Y\ \big|\ G(\xi,w)=0 \ \text{for all } w\in TY\bigr\}\,.
\end{equation}
As usual, one says that $Y$ is:
\begin{gather}
\begin{aligned}\label{eq:defs_isotropy}
&\text{isotropic:}\qquad  TY\subseteq (TY)^\perp\,,\\
&\text{coisotropic:}\quad   (TY)^\perp\subset TY\,,\\
&\text{Lagrangian:}\quad  TY=(TY)^\perp\,.
\end{aligned}
\end{gather}
For a positive-definite metric these notions are essentially trivial, since $TY\cap (TY)^\perp=0$. For the neutral metric \eqref{eq:neutral_metric}, however, the inclusion
\begin{equation}
(TY)^\perp\subset TY
\end{equation}
can hold nontrivially, and this is precisely the coisotropic situation relevant for topological defects.

To connect this geometry with the worldsheet conditions, choose local Lorentzian coordinates $(\sigma^0,\sigma^1)$ such that the defect lies at $\sigma^1=0$ and $\sigma^0$ is oriented along it. Using the defect map $\Phi_D$ defined in \eqref{eq:target-space-map}, set
\begin{equation}\label{eq:defect-tangent-normal-vectors}
u=\left.\bigl(\partial_0\phi_L,\partial_0\phi_R\bigr)\right|_D\,,\quad v=\left.\bigl(\partial_1\phi_L,\partial_1\phi_R\bigr)\right|_D\,.
\end{equation}
Since $\Phi_D(D)\subset Y$, its tangential derivative satisfies $u\in TY$, while $v$ collects the one-sided normal derivatives of the two bulk fields. This is illustrated schematically in Figure \ref{fig:tangent_space_Y}.

The translation of the energy-momentum tensor gluing condition to target-space geometry constrains derived in \cite{Kapustin:2010zc}, can be summarise in two level. A (correspondence space) metric formulation and a more intrinsic formulation in terms of the geometry. We will use mostly the latter in the example, but it is useful to first introduce the former.

%%%%%%%%%
\paragraph{Metric geometric formulation.}
A pair
\begin{equation}
u\in TY\,, \quad v\in T(X_L\times X_R)|_Y
\end{equation}
is called allowed when it satisfies the action-derived gluing condition
\begin{equation}\label{eq:KS_allowed_pair}
G(v,w)=F(u,w) \quad \text{for all }w\in TY\,.
\end{equation}
Then, in terms of this geometric language, the conformal defect condition \eqref{eq:conformal-topological-conditions}, becomes 
\begin{equation}
T^1{}_0-\widehat T^1{}_0=0 \quad\Longleftrightarrow\quad G(u,v)=0\,.
\end{equation}
Note that this is automatic for any allowed pair, since setting $w=u$ in \eqref{eq:KS_allowed_pair} gives
\begin{equation}
G(u,v)=G(v,u)=F(u,u)=0\,,
\end{equation}
by antisymmetry of $F$. Thus, within the class of defect actions defined in Section \ref{sec:WZW-models-defects}, the action-derived gluing condition automatically satisfies the classical conformal defect condition.

Topologicality however imposes additional requirements. In particular, the authors in \cite{Kapustin:2010zc}  prove that the topological stress-tensor condition holds for every allowed pair if and only if every allowed pair $(u,v)$ also satisfies $v\in TY$ and the exchanged pair $(v,u)$ is allowed. With the convention \eqref{eq:KS_allowed_pair}, the latter condition is
\begin{equation}\label{eq:KS_topological_pair}
G(u,w)=F(v,w) \quad \text{for all }w\in TY\,.
\end{equation}

%%%%%%%%
\paragraph{Intrinsic geometric formulation.}
The allowed-pair criterion admits an equivalent formulation entirely in terms of the geometry of $(Y,F)$. First, topologicality requires the support to be coisotropic with respect to the neutral metric $G$, i.e $(TY)^\perp\subset TY$. Choose a complementary subbundle $S_Y\subset TY$ such that $TY=(TY)^\perp\oplus S_Y$. This is called a screen distribution in \cite{Kapustin:2010zc}. It is naturally isomorphic to the quotient
\begin{equation}\label{eq:screen-quotient}
\mathcal Q_Y:=TY/(TY)^\perp\,.
\end{equation}
We denote the restrictions of $G$ and $F$ to $S_Y$ by
\begin{equation}
\widetilde G:=G|_{S_Y}\,, \quad\widetilde F:=F|_{S_Y}\,.
\end{equation}
% Equivalently, these are the tensors induced on $\mathcal Q_Y$. The metric $\widetilde G$ is nondegenerate, while $\widetilde F$ is nondegenerate precisely when
% \begin{equation}
% \ker F=(TY)^\perp\,.
% \end{equation}

%%%%%%%%
\paragraph{Topologicality criterion.}
Kapustin and Setter prove that the topological defect condition is equivalent to
\begin{equation}\label{eq:KS_geometric_conditions}
(TY)^\perp\subset TY\,, \quad \ker F=(TY)^\perp\,, \quad \bigl(\widetilde G^{-1}\widetilde F\bigr)^2 = \mathrm{id}_{\mathcal Q_Y}\,.
\end{equation}
Thus one checks coisotropy of the support, equality of the null directions of $G|_{TY}$ and $F$, and involutivity on the quotient $\mathcal Q_Y$.

\medskip
Finally let us make an important comment for what is to follow. The allowed-pair relation and the intrinsic topologicality criterion can in fact also be used constructively. Indeed, given two target geometries and a symmetry-motivated candidate support $Y\subset X_L\times X_R$, one may begin with an ansatz for $F$. Equation \eqref{eq:KS_allowed_pair} determines the associated gluing condition, whose conformality is automatic, while \eqref{eq:KS_geometric_conditions} selects the topological candidates, refining the starting Ansatz. These candidates must also satisfy \eqref{eq:bibrane-curvature} and one can then determine compatible global connection data, effectively deriving the defect action as in \eqref{eq:defect-action}. We use this strategy in Section \ref{sec:constructions} to construct or verify the defects considered there.

%%%%%%%%%%%%%%%%%%%%%%%%%%%%%%%%%%%%%%%%%%%%%%%%
%%%%%%%%%%%%%  Classical fusion   %%%%%%%%%%%%%%
%%%%%%%%%%%%%%%%%%%%%%%%%%%%%%%%%%%%%%%%%%%%%%%%
\section{Classical fusion through geometry}
\label{sec:classical_fusion}

We now turn to the central construction of this paper. The target-space composition of bi-branes was discussed in \cite{Fuchs:2007fw} in terms of a fibre product over the intermediate target followed by projection onto the external targets, together with the corresponding bundle data. For explicit sigma-model calculations, however, one still needs a criterion determining when the combined two-form and connection data descend through this projection and whether descent holds on the whole correspondence or only on particular loci. We formulate these requirements below as local curvature and global holonomy conditions.

Starting from two composable topological defects
\begin{equation}
(Y_{LI},F_{LI},\nabla_{LI}) \quad\text{and}\quad (Y_{IR},F_{IR},\nabla_{IR})\,,
\end{equation}
the construction produces a labelled collection of descended defects between $X_L$ and $X_R$. Only those branches that also satisfy the topologicality conditions of Section \ref{sec:review} are retained in the fusion product. This makes the possible branch structure an output of the target-space reduction, rather than something that must be inferred by disentangling the summed worldsheet action. Conversely, the local one-form entering the action of each resulting branch can be reconstructed from its descended connection data through \eqref{eq:defect-action}.

The construction is summarised schematically by
\begin{equation}\label{eq:geometric-fusion-diagram}
\raisebox{3pt}{\RawGeometricFusionDiagram} \;\;\;\raisebox{1pt}{\GeometricFusionArrow}\hspace{3pt}\;\;\;\raisebox{3pt}{\FusedGeometricFusionDiagram}\hspace{2pt}\,.
\end{equation}
On the right-hand side, we have suppressed the corresponding connection-descent relation $\pi_a^*\nabla_{LR}^{(a)}\cong\nabla_{LR}^{\rm raw}|_{Z_a}$ and $a\in\mathcal A$ labels the admissible loci $Z_a$.

The prescription we lay out consists of three steps. First, we form the raw composition of the two supports and combine their two-form and connection data. Second, we eliminate the intermediate target variable by restricting to the loci on which these data descend through the projection onto $X_L\times X_R$. Third, we retain the descended branches that satisfy the topologicality criterion \eqref{eq:KS_geometric_conditions}.\footnote{If the final step is omitted, the procedure still produces descended defect data. The gluing conditions derived from their local defect actions are conformal as explained in Section \ref{subsec:defect conf-topo-conditions}; topologicality is the additional requirement imposed here.} Let us turn to  describe these steps in detail.

%%%%%%%%%
\paragraph{Raw composition.}
Consider two composable defects
\begin{equation}
	(Y_{LI},F_{LI},\nabla_{LI})\,, \quad (Y_{IR},F_{IR},\nabla_{IR})\,,
\end{equation}
where for each the global connection data understood as in Section \ref{sec:review}. Composability means that the right-hand background of the first defect agrees with the left-hand background of the second, including its Wess-Zumino data. Their intrinsic two-forms satisfy
\begin{equation}
dF_{LI}=p_L^*H_L-p_I^*H_I\,, \quad dF_{IR}=p_I^*H_I-p_R^*H_R\,.
\end{equation}
Locally, the connections are represented by $A_{LI}$ and $A_{IR}$ through the usual relation $F_{LI}=p_L^*b_L-p_I^*b_I+dA_{LI}  $.

Let $(Y_{LI},F_{LI},\nabla_{LI})$ and $(Y_{IR},F_{IR},\nabla_{IR})$ be composable defects. Their raw correspondence is the fibre product
\begin{equation}\label{eq:raw-fibre-product}
Y_{LR}^{\rm raw} = Y_{LI}\times_{X_I}Y_{IR} = \left\{ (x_L,x_I,x_R) \ \middle|\ (x_L,x_I)\in Y_{LI},\ (x_I,x_R)\in Y_{IR} \right\}\,.
\end{equation}
Denoting the associated projections by
\begin{equation}
\mathrm{pr}_{LI}:Y_{LR}^{\rm raw}\longrightarrow Y_{LI}\,, \quad \mathrm{pr}_{IR}:Y_{LR}^{\rm raw}\longrightarrow Y_{IR}\,,
\end{equation}
the raw connection and intrinsic two-form are
\begin{align}
\nabla_{LR}^{\rm raw} &= \mathrm{pr}_{LI}^*\nabla_{LI} \otimes \mathrm{pr}_{IR}^*\nabla_{IR}\,, \label{eq:fusion-raw-connection} \\
F_{LR}^{\rm raw} &= \mathrm{pr}_{LI}^*F_{LI} + \mathrm{pr}_{IR}^*F_{IR}\,. \label{eq:raw-intrinsic-form}
\end{align}
The tensor product in \eqref{eq:fusion-raw-connection} denotes the composition of the possibly twisted connection data and composability ensures that their intermediate twists cancel. In compatible local trivialisations, the corresponding one-form is $A_{LR}^{\rm raw} = \mathrm{pr}_{LI}^*A_{LI} + \mathrm{pr}_{IR}^*A_{IR}$. The intermediate bulk potential cancels, yielding
\begin{equation}\label{eq:raw-local-connection-relation}
F_{LR}^{\rm raw} = p_L^*b_L-p_R^*b_R+dA_{LR}^{\rm raw}\,.
\end{equation}

The raw correspondence projects onto the external targets through
\begin{equation}\label{eq:proj-raw}
\pi_{LR}:Y_{LR}^{\rm raw}\longrightarrow X_L\times X_R\,, \qquad \pi_{LR}(x_L,x_I,x_R)=(x_L,x_R).
\end{equation}
Its projected image determines the possible fused support, but the differential data must still descend through the factorisation fibres of $\pi_{LR}$, to which we now turn

%%%%%%%%%
\paragraph{Descent.}  

While the projection $\pi_{LR}$ eliminates the intermediate field. For fixed external data $(x_L,x_R)$, there may be several compatible values of this intermediate field. Altogether, these possible values, form the corresponding factorisation fibre. For the raw data we just constructed to define a defect depending only on $(x_L,x_R)$, they must be insensitive or independent to motion within this fibre. Locally, this requires the raw curvature to have no component along the fibre directions. Globally, the raw connection must have trivial holonomy around closed fibre cycles. If a fibre has several connected components, the specific data that characterise each component has to be determined. We now describe how to impose these conditions using the target-space geometry.

Let $Z\subset Y_{LR}^{\rm raw}$ be a smooth locus such that the restricted map
\begin{equation}
\pi_{LR}|_Z: Z\longrightarrow Y_{LR}:=\pi_{LR}(Z)
\end{equation}
is a surjective submersion with embedded image. In addition we demand that the selected locus contains connected components of the original factorisation fibres. Hence, in what follows the reduction strata used in the examples will be specified separately. Now, the vectors tangent to the factorisation fibres are the vertical vectors
\begin{equation}
V\in\ker d(\pi_{LR}|_Z)\,.
\end{equation}
The local curvature condition is
\begin{equation}\label{eq:fusion-cond-horiz}
\iota_V\!\left(F_{LR}^{\rm raw}|_Z\right)=0 \quad \text{for every } V\in\ker d(\pi_{LR}|_Z)\,.
\end{equation}
This also implies invariance along the fibres. Indeed,
\begin{equation}
dF_{LR}^{\rm raw}=p_L^*H_L-p_R^*H_R
\end{equation}
is pulled back from the external targets, and hence Cartan's formula gives $\mathcal L_VF_{LR}^{\rm raw} =d\iota_VF_{LR}^{\rm raw}+\iota_VdF_{LR}^{\rm raw}=0$. Thus $F_{LR}^{\rm raw}|_Z$ is basic\footnote{A differential form $\omega\in\Omega^\bullet(Z)$ is basic with respect to a surjective submersion $\pi:Z\to Y$ if it is horizontal and invariant along the fibres: 
\begin{equation}
	\iota_V\omega=0\,, \quad \mathcal L_V\omega=0
\end{equation}
for every vertical vector field $V\in\ker d\pi$. Equivalently, $\omega=\pi^*\widetilde\omega$ for a unique form $\widetilde\omega$ on $Y$.} and determines a unique two-form $F_{LR}$ on $Y_{LR}$ satisfying
\begin{equation}\label{eq:fused-form-descent}
(\pi_{LR}|_Z)^*F_{LR}=F_{LR}^{\rm raw}|_Z\,.
\end{equation}

Because the external bulk potentials in \eqref{eq:raw-local-connection-relation} are pulled back from $X_L\times X_R$, for every vertical vector field $V$ one has
\begin{equation}
\iota_VdA_{LR}^{\rm raw} = \iota_VF_{LR}^{\rm raw}\,.
\end{equation}

The curvature condition \eqref{eq:fusion-cond-horiz}  does not by itself ensure descent of the connection. Its restriction to each factorisation fibre is flat, but may have non-trivial holonomy. Global descent therefore additionally requires
\begin{equation}\label{eq:fibre-holonomy}
\mathrm{Hol}_{\nabla_{LR}^{\rm raw}} (\gamma_{\rm fib})=1
\end{equation}
for every closed loop $\gamma_{\rm fib}$ contained in a fibre of $\pi_{LR}|_Z$. When the fibres are disconnected, trivial holonomy controls the connection only on each connected component. Descent additionally requires compatible identifications between the connection data over the different components, satisfying the usual cocycle condition.  In a trivialisation along the loop, this condition is represented by
\begin{equation}
\exp\left( i\oint_{\gamma_{\rm fib}}A_{LR}^{\rm raw} \right)=1\,.
\end{equation}
For connected factorisation fibres, equations \eqref{eq:fusion-cond-horiz} and \eqref{eq:fibre-holonomy} are the local and global conditions for connection descent. The raw connection therefore descends to a connection $\nabla_{LR}$ on $Y_{LR}$.

%%%%--------- figure ---------%%%%
\begin{figure}[t]
    \centering
    \includegraphics[width=0.65\linewidth]{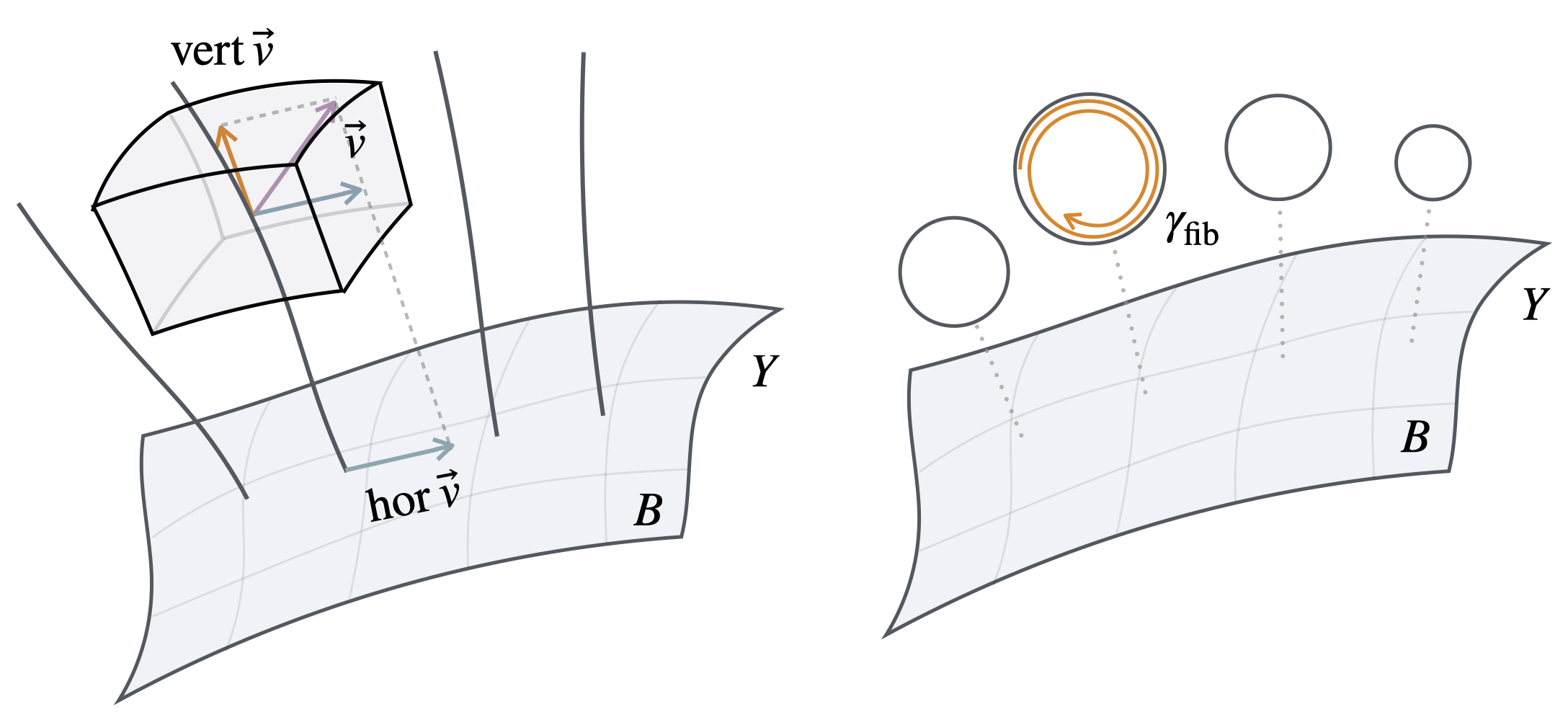}
    \caption{Local and global descent conditions for $\pi_a:Z_a\to Y_{LR}^{(a)}$. Left: the curvature-horizontality condition \eqref{eq:fusion-cond-horiz} requires $F_{LR}^{\rm raw}(V,W)=0$ for every $V\in\ker d\pi_a$ and $W\in T_zZ_a$. Right: the global condition \eqref{eq:fibre-holonomy} requires $\mathrm{Hol}_{\nabla_{LR}^{\rm raw}}(\gamma_{\rm fib})=1$ for every closed loop $\gamma_{\rm fib}\subset\pi_a^{-1}(y)$.}
    \label{fig:descent}
\end{figure}
%%%%--------- figure ---------%%%%

%%%%%%%%
\paragraph{Fusion branches.}

The descent conditions need not hold on the whole raw correspondence. They may instead select several connected loci, each of which produces a separate candidate defect after projection. We will call the resulting defects the fusion branches. The different labels then distinguishes the different loci they may project to, as they may have the same support while carrying different two-form or connection data. This is the classical counterpart of the object label in categorical fusion, here however classical defects are characterised by their target space support as well as the two-form it carries.

We call a smooth locus $Z\subset Y_{LR}^{\rm raw}$ descent-admissible when the restricted projection satisfies the regularity, curvature-descent and fibre-holonomy conditions above, together with the required compatibility between fibre components when the fibres are disconnected. We move on to choose connected regular reduction strata $Z_a$ and subsequently test the descent procedure on each of them. Writing
\begin{equation}
Y_{LR}^{(a)}:=\pi_{LR}(Z_a)\,,\qquad \pi_a:=\pi_{LR}|_{Z_a}: Z_a\longrightarrow Y_{LR}^{(a)}\,,
\end{equation}
the raw data descend on every such stratum to
\begin{equation}\label{eq:admissible-branch-descent}
\bigl( Y_{LR}^{(a)}, F_{LR}^{(a)}, \nabla_{LR}^{(a)} \bigr)\,, \quad
\pi_a^*F_{LR}^{(a)} = F_{LR}^{\rm raw}|_{Z_a}\,.
\end{equation}
Descent alone does not guarantee that the resulting branch is topological. For the fusion of topological defects considered here, we retain only those descended branches for which $(Y_{LR}^{(a)},F_{LR}^{(a)})$ satisfies the topologicality criterion \eqref{eq:KS_geometric_conditions}. We call these branches fusion-admissible and denote their label set by $\mathcal A$. 

The classical fusion prescription therefore assigns to two composable defects the labelled collection of their admissible descended branches:
\begin{equation}\label{eq:classical-fusion-prescription}
(Y_{LI},F_{LI},\nabla_{LI}) \star (Y_{IR},F_{IR},\nabla_{IR}) := \left\{ Y_{LR}^{(a)}, F_{LR}^{(a)}, \nabla_{LR}^{(a)} \right\}_{a\in\mathcal A}\,.
\end{equation}
Thus $\star$ denotes the following sequence: from the fibre product \eqref{eq:raw-fibre-product}, combine the connection and curvature data according to \eqref{eq:fusion-raw-connection} and \eqref{eq:raw-intrinsic-form}, project by \eqref{eq:proj-raw}, restrict to the admissible loci selected by \eqref{eq:fusion-cond-horiz} and \eqref{eq:fibre-holonomy}, and descend the resulting data through \eqref{eq:admissible-branch-descent}.\footnote{We use this prescription for topological defects. For generic conformal defects, bringing the defect lines together may require a short-distance regularisation and can generate an defect renormalisation-group flow \cite{Bachas:2007td}.}

The underlying fused support is
\begin{equation}\label{eq:total-fused-support}
Y_{LR}^{\rm fus} = \bigcup_{a\in\mathcal A}Y_{LR}^{(a)}\,.
\end{equation}
This set-theoretic union does not contain the complete branch data. If the supports are mutually disjoint, the indexed family may equivalently be written as
\begin{equation}
(Y_{LI},F_{LI},\nabla_{LI}) \star (Y_{IR},F_{IR},\nabla_{IR}) \equiv \bigsqcup_{a\in\mathcal A} \bigl( Y_{LR}^{(a)}, F_{LR}^{(a)}, \nabla_{LR}^{(a)} \bigr)\,.
\end{equation}
When branches meet, their labels and differential data must instead be retained separately; their union need not carry a single smooth defect structure.

The two parts of the construction should therefore be distinguished: \eqref{eq:fusion-cond-horiz} and \eqref{eq:fibre-holonomy} determine whether the composite connection data descend, while \eqref{eq:KS_geometric_conditions} determines whether the descended branch is topological. Conformality of the action-derived gluing condition follows from Section \ref{subsec:defect conf-topo-conditions}. The corresponding conditions are summarised in Table \ref{tab:defect-conditions-summary}.

Lastly, we will at times also loosen slightly the definition of target-space fusion given here. When only the support and intrinsic
two-form are reduced: regularity, curvature descent and pointwise
topologicality are imposed, but global connection descent is not, we fill denote the associate composition by $\star_{\mathrm{loc}}$. The result then define a locally define pair $(Y,F)$ (i.e. contrary to the operation denoted above by  $\star$, the full fusion product carrying the descended connection data $\nabla$.

%%%----------- Condition table ---------%%%%
\begin{table}[t]
\centering
\small
\setlength{\tabcolsep}{6pt}
\renewcommand{\arraystretch}{1.55}
\begin{tabularx}{\textwidth}{
    @{}
    >{\raggedright\arraybackslash}p{0.15\textwidth}|
    >{\raggedright\arraybackslash}p{0.39\textwidth}
    >{\raggedright\arraybackslash}X
    @{}
}
\toprule
\cellcolor{black!3} Level
&
\cellcolor{black!3} Conformal defect condition
&
\cellcolor{black!3} Additional topological condition
\\
\midrule

\cellcolor{black!3} Stress tensor
&
$\displaystyle
T^1{}_0-\widehat T^1{}_0=0$
&
$\displaystyle
% T^1{}_0-\widehat T^1{}_0=0\,,\quad
T^1{}_1-\widehat T^1{}_1=0$
\\

\cellcolor{black!3} $\begin{array}{@{}l@{}}
\text{Allowed pairs}\\[-8pt]
\forall\,w\in TY
\end{array}$
& $\displaystyle
\begin{array}{@{}l@{}}
G(v,w)=F(u,w)\,,\quad u\in TY%\\[-10pt]
% G(u,w)=F(v,w)\,,\quad v\in TY
\end{array}$
&
$\displaystyle
G(u,w)=F(v,w)\,,\quad v\in TY$
\\

\cellcolor{black!3} Geometry of the support
&
No additional condition beyond the
action-derived gluing relation
&
$\displaystyle
(TY)^\perp\subset TY\,,\quad
\ker F=(TY)^\perp$
\\

\cellcolor{black!3} Quotient bundle
&
$\qquad\qquad\qquad$---
&
$\displaystyle
\bigl(\widetilde G^{-1}\widetilde F\bigr)^2
=\mathrm{id}_{\mathcal Q_Y}$
\\

\cellcolor{black!3} Fusion
&
Not defined here in general; fusion may require
short-distance regularisation
&
Connection-level descent followed by
\eqref{eq:KS_geometric_conditions}
\\

\bottomrule
\end{tabularx}
\caption{Equivalent formulations of the classical conformal and
topological defect conditions, and their role after fusion.}
\label{tab:defect-conditions-summary}
\end{table}
%%%--------------------------------------------%%%%

%%%%%%%%%%%%%%%%%%%%%%%%%%%%%%%%%%%%%%%%%%%%%%%%%%%%%%
%%%%%%%%%%%%      Defects and fusion      %%%%%%%%%%%%
%%%%%%%%%%%%%%%%%%%%%%%%%%%%%%%%%%%%%%%%%%%%%%%%%%%%%%
\section{Construction of defects and their classical fusion}
\label{sec:constructions}

In this section we will now apply the target-space framework laid out in the previous section. Specifically, we use it in two complementary ways. First, starting from two sigma-models and a symmetry-motivated ansatz for the defect support, the geometric topologicality conditions constrain the intrinsic two-form carried by the defect. Hence using this target-space formulation of defects, this provides a systematic way of constructing and testing candidate topological defects. Second, once the defect data are known, the prescription of Section \ref{sec:classical_fusion} determines their classical fusion.

As highlighted earlier, although we adopt the more convenient target space formulation, the worldsheet action description of the defect is not lost. On local bulk patches, the intrinsic data $(Y,F)$ determine a defect one-form $A$ through
\begin{equation}\label{eq:two-form-F-intro-sec4}
F=p_L^*b_L-p_R^*b_R+dA\,,
\end{equation}
so that $\int_\ell A$ gives the corresponding local coupling in the worldsheet action. Globally however, this one-form is replaced by the corresponding twisted connection datum. The target-space description therefore contains the local action data while providing the additional geometric information needed to identify the branches produced by fusion.

That is we construct and consider topological defects $(Y,F)$ that can be placed between two sigma-modes with target-spaces $X_{L/R}$ and associated geometric data, schematically 
\begin{equation}\label{eq:construction-fusion-recipe}
\raisebox{3pt}{\DefectConstructionDiagram}\,. %\qquad \xrightarrow{\;\text{fusion and descent}\;} \qquad \bigsqcup_{a\in\mathcal A} \bigl(Y_{LR}^{(a)},F_{LR}^{(a)},\nabla_{LR}^{(a)}\bigr)\,.
\end{equation}
We consider pairs $(Y,F)$ satisfying the curvature identity \eqref{eq:bibrane-curvature} and the topologicality criterion \eqref{eq:KS_geometric_conditions}, and then apply the fusion prescription of Section \ref{sec:classical_fusion}.

We begin with the standard WZW bi-branes and recover their classical biconjugacy branches through descent of the intrinsic two-form. We then compose the degree-$p$ Hopf defect with its orientation reverse. The admissible loci form a finite family of fibre-shift graph defects selected by the holonomy of the raw connection around the intermediate circle. We subsequently construct T-duality--shift--T-duality (TsT)  defects by inserting a shear in the T-dual frame. Finally, we compose a WZW bi-brane with the WZW-compatible Hopf defect. The resulting T-dual biconjugacy correspondence illustrates the distinction between regular branchwise descent and singular non-clean loci. The latter two defects (the TsT defect and T-dual bibrane) have, to our knowledge, not yet been derived as classical defects in the literature, and they demonstrate the constructive power of the fusion operation defined in Section \ref{sec:classical_fusion}.  The resulting classical fusion rules are summarised in Table \ref{tab:classical-defect-fusion}. 

Throughout this section we at times suppress the global connection datum $\nabla$ from the notation unless it is essential. Whenever the full fusion product $\star$ is used, the descended connection data are understood as in Section \ref{sec:classical_fusion}.

%%%%%%%%%%%%%%%%%%%%%%%%%%%%%%%%%%%%%%%%%%%%%%%%
%%%%%%%%%%%%%%%%%%%%%%%%%%%%%%%%%%%%%%%%%%%%%%%%
\subsection{WZW bi-branes} \label{subsec:classical_fusion_bibranes}

We now turn to the symmetry-preserving WZW bi-branes of \cite{Fuchs:2007fw}. These class of defects are the analogues of the much more familiar description of symmetry-preserving WZW boundary conditions: the corresponding D-brane worldvolumes are conjugacy classes in the group manifold (or also twisted) conjugacy classes for automorphism-twisted gluing conditions \cite{Klimcik:1996hp,Alekseev:1998mc,Stanciu:1999id}. For a defect, the two worldsheet fields take values in $G\times G$, and the conjugacy-class condition is instead imposed on their relative group element.

Accordingly, for a compact Lie group $G$ bi-brane defects are given by so-called biconjugacy classes, namely preimages of ordinary conjugacy classes under the relative multiplication map\footnote{The map denoted by $\mu$ here is denoted by $\widetilde\mu$ in \cite{Fuchs:2007fw}.}
\begin{equation}\label{eq:mu-tilde-SU2}
\mu:SU(2)\times SU(2)\to SU(2)\,, \qquad \mu(g_1,g_2)=g_1g_2^{-1}\,,
\end{equation}
where we specialise for further ease of exposition and computations to $G=SU(2)$.
The conjugacy classes of $SU(2)$ are labelled by a class angle $\alpha\in[0,\pi]$. For $\alpha\in(0,\pi)$ they are two-spheres, while $C_0=\{\mathbf 1\}$ and $C_\pi=\{-\mathbf 1\}$. Hence one can parametrise the symmetry-preserving defect worldvolumes as a one-parameter family 
\begin{equation}\label{eq:Balpha-def}
B_\alpha = \mu^{-1}(C_\alpha) = \left\{ (g_1,g_2)\in SU(2)\times SU(2) \ \middle|\ g_1g_2^{-1}\in C_\alpha \right\}\,.
\end{equation}
Thus $B_\alpha$ realises the general defect support $Y_{ij}$ for $X_i=X_j=SU(2)$. Within the standard symmetry-preserving family, the label $\alpha$ determines the biconjugacy-class support and its associated intrinsic two-form. The associated intrinsic two-form may be written as
\begin{equation}
\label{eq:Falpha-standard}
F_\alpha = \mu^*\omega_\alpha -\frac{k}{2}\,\langle p_1^*\theta^L\wedge p_2^*\theta^L\rangle\,,
\end{equation}
where $\theta^L=g^{-1}dg$ is the left Maurer-Cartan form, $p_1,p_2$ are the two projections from $SU(2)\times SU(2)$, and $\omega_\alpha$ is the canonical two-form on the conjugacy class $C_\alpha$. Note that strico-senso the first term in \eqref{eq:Falpha-standard} is the pullback of $\omega_\alpha$ along $\mu|_{B_\alpha}:B_\alpha\to C_\alpha$ and in a slight abuse of notation we will leave the restriction implicit. By construction,
\begin{equation}\label{eq:dFalpha-fusion}
dF_\alpha=p_1^*H-p_2^*H\,.
\end{equation}
The data $(B_\alpha,F_\alpha)$ are precisely the target-space data of the symmetry-preserving topological WZW defects identified in \cite{Fuchs:2007fw}. In particular, they satisfy the topological defect gluing conditions. These WZW bi-branes also provide a standard class of examples for the general target-space characterisation of topological sigma-model defects of \cite{Kapustin:2010zc}.

These bi-branes (or rather those that verify (pre-)quantisation conditions) are the semiclassical target-space realisations of  symmetry-preserving Verlinde defects of the (diagonal) WZW model. We will return below to this relation between their classical composition and the well-known Verlinde fusion rules. When considering the global connection, we always take $k \in \mathbb Z>0$ and $\alpha=\pi a/k\,$ with $a=0,\dots, k$, while the continuous-angle formulas describe local (classical) data.

For later use, let us also detail a convenient $SU(2)$ parametrisation. Using Hopf coordinates, a group element of $g\in SU(2)$ can be written as
\begin{equation}\label{eq:SU2-Hopf-param}
g(\theta,\psi,\phi) =
\begin{pmatrix}
e^{\,i(\phi-\psi/2)}\cos\frac{\theta}{2} & -\,e^{-i(\phi+\psi/2)}\sin\frac{\theta}{2} \\[2mm]
e^{\,i(\phi+\psi/2)}\sin\frac{\theta}{2} & e^{-i(\phi-\psi/2)}\cos\frac{\theta}{2}
\end{pmatrix}\,,
\end{equation}
one has the defining $g^\dagger g=\mathbf 1$ and $\det g=1$, and this parametrisation leads to the Hopf-fibration metric. Given this parametrisation, the defining constraint of the $SU(2)$-bi-brane \eqref{eq:Balpha-def} biconjugacy may be written explicitly  as
\begin{align}\label{eq:Balpha-Hopf-eq}
\cos\alpha =& \frac{1}{2}\mathrm{tr}(g_1g_2^{-1}) \nonumber\\
=&\cos(\phi_1-\phi_2)\, \cos\tfrac{\theta_1-\theta_2}{2}\, \cos\tfrac{\psi_1-\psi_2}{2} + \sin(\phi_1-\phi_2)\, \cos\tfrac{\theta_1+\theta_2}{2}\, \sin\tfrac{\psi_1-\psi_2}{2}\,.
\end{align}
Thus, for generic $\alpha\in(0,\pi)$, $B_\alpha$ is a five-dimensional submanifold of $SU(2)\times SU(2)$.

For later use it is useful to record two special instances. The degenerate cases $\alpha=0,\pi$ reduce to defects imposing the gluing conditions
\begin{equation}\label{eq:special-case-bibranes}
B_0=\{g_1=g_2\}\,, \qquad B_\pi=\{g_1=-g_2\}\,.
\end{equation}
These are the graph defects associated with the identity map and with multiplication by the non-trivial central element $-\mathbf 1$, respectively.

%%%%%%%%%%%%%%%%%%%%%%%%%%%%%%%%%%%%%%%%%%%%%%%%
%%%%%%%%%%%%%%%%%%%%%%%%%%%%%%%%%%%%%%%%%%%%%%%%
\subsubsection{Classical fusion of $SU(2)$ bi-branes}

We now use the prescription we defined in Section \ref{sec:classical_fusion} to determine the classical composition of two symmetry-preserving $SU(2)$ bi-branes,
\begin{equation}\label{eq:fusion-bibranes}
(B_\alpha,F_\alpha)\star(B_\beta,F_\beta)\,.
\end{equation}
The fusion of WZW bi-branes was already discussed and partly derived (at the level of the support) in \cite{Fuchs:2007fw}. In particular, the classical product of conjugacy classes was shown to reproduce the allowed range of fusion channels. The authors noted that geometric quantisation would relate those classical ``fusion channels'' to the $\widehat{\mathfrak g}_k$-fusion rules. Our purpose here is to express this standard example in the descent formulation used throughout this paper: on each regular branch, we show explicitly that the raw two-form descends to the standard bi-brane two-form. This calculation will serve as a benchmark for the Hopf and mixed fusions considered below.

%%%%%%
\paragraph{Fused support.}
At the level of support, the raw composition is the fibre product over the middle copy of
$SU(2)$. Specialising the fibre product of
Section \ref{sec:classical_fusion} to the present case gives
\begin{equation}\label{eq:raw-bibrane-fusion-support}
Y_{\alpha\beta}^{\mathrm{raw}} := B_\alpha\times_{SU(2)}B_\beta = \left\{ (g_1,g_2,g_3)\in SU(2)^3 \ \middle|\ g_1g_2^{-1}\in C_\alpha,\; g_2g_3^{-1}\in C_\beta \right\}\,.
\end{equation}
Projecting to the first and third factors gives
\begin{equation}
Y_{\alpha\beta}^{\mathrm{eff}} := \pi_{13}\bigl(Y_{\alpha\beta}^{\mathrm{raw}}\bigr) \subset SU(2)\times SU(2)\,.
\end{equation}
Using the defining constraints in \eqref{eq:raw-bibrane-fusion-support}, one finds
immediately $g_1g_3^{-1} = (g_1g_2^{-1})(g_2g_3^{-1}) \in C_\alpha C_\beta$, hence
\begin{equation}\label{eq:bibrane-support-product}
Y_{\alpha\beta}^{\mathrm{eff}} = \mu^{-1}(C_\alpha C_\beta)\,, \qquad \mu(g_1,g_3)=g_1g_3^{-1}.
\end{equation}
Thus the support-level fusion problem reduces to the product of conjugacy classes in $SU(2)$. Every group element may be written as
\begin{equation}\label{eq:galpha-n}
g_\alpha(\mathbf n) = \cos\alpha\,\mathbf 1 +i\sin\alpha\,\mathbf n\cdot\sigma\,, \qquad \alpha\in[0,\pi]\,, \quad \mathbf n\in S^2.
\end{equation}
Geometrically, this realises $SU(2)\simeq S^3$ as a family of two-spheres parametrised by $\alpha$, with the sphere collapsing to the central elements $\mathbf 1$ and $-\mathbf 1$ at $\alpha=0$ and $\alpha=\pi$, respectively. Conjugation rotates $\mathbf n$ while leaving $\alpha$ fixed. Hence, for $0<\alpha<\pi$, the conjugacy class $C_\alpha$ is the two-sphere obtained by varying $\mathbf n$, while
\begin{equation}
C_0=\{\mathbf 1\}\,, \qquad C_\pi=\{-\mathbf 1\}\,.
\end{equation}

Writing an element of $C_\beta$ similarly, one finds
\begin{equation}
\frac12\, \mathrm{tr}\!\left( g_\alpha(\mathbf n)g_\beta(\mathbf m) \right) = \cos\alpha\cos\beta -\sin\alpha\sin\beta\,\mathbf n\cdot\mathbf m\,.
\end{equation}
Since the conjugacy class of an element $g_\gamma$ is determined by $\frac12\mathrm{tr}(g_\gamma)=\cos\gamma$, varying $\mathbf n\cdot\mathbf m\in[-1,1]$ determines the possible values of $\gamma$ in the product $C_\alpha C_\beta$.  When parametrising the associated conjugacy class as in \eqref{eq:Balpha-Hopf-eq}, the possible angles $\gamma$ are similarly constrained to an interval 
\begin{equation}\label{eq:gamma-interval}
\gamma\in I_{\alpha,\beta}\,, \qquad I_{\alpha,\beta} = \left[ |\alpha-\beta|, \min(\alpha+\beta,\,2\pi-\alpha-\beta) \right]\,.
\end{equation}
Substituting into \eqref{eq:bibrane-support-product}, we obtain
\begin{equation}\label{eq:bibrane-fusion-support}
Y_{\alpha\beta}^{\mathrm{eff}} = \bigcup_{\gamma\in I_{\alpha,\beta}}B_\gamma\,,
\end{equation}
Thus the projected support decomposes into the disjoint biconjugacy-class level sets $B_\gamma$, with $\gamma\in I_{\alpha,\beta}$. This support-level composition agrees with the description of WZW bi-brane fusion in \cite{Fuchs:2007fw}. However, this is only ``half'' the data of characterising the topological defect at the level of the target-space; whether the differential data descend to these level sets must still be checked.

%%%%%%
\paragraph{Fused two-form.}
 We now turn to how the two-form data of the defect behave under the same composition. We use the bi-brane two-form defined in \eqref{eq:Falpha-standard}, which satisfies \eqref{eq:dFalpha-fusion}. For the composition of $B_\alpha$ and $B_\beta$, the raw two-form is $F_{\alpha\beta}^{\rm raw} = {\rm pr}_{12}^*F_\alpha + {\rm pr}_{23}^*F_\beta$. With this orientation convention the intermediate $H$-flux cancels:
\begin{equation}
dF_{\alpha\beta}^{\rm raw} = p_1^*H-p_3^*H\,.
\end{equation}
 Now in order to parametrise the different conjugacy classes efficiently, let
\begin{equation}\label{eq:param-bibrane-fusion}
q_{12}:=g_1g_2^{-1}\in C_\alpha\,, \qquad q_{23}:=g_2g_3^{-1}\in C_\beta\,, \qquad r:=g_1g_3^{-1}=q_{12}q_{23}\,,
\end{equation}
Fix $\gamma$ in the interior of $I_{\alpha,\beta}$ and work on $\pi_{13}^{-1}(B_\gamma)$, where $r\in C_\gamma$. We again denote the restricted projection to $B_\gamma$ by $\pi_{13}$. For fixed $(g_1,g_3)\in B_\gamma$, its fibre is
\begin{equation}\label{eq:clean-locus-bibrane}
\pi_{13}^{-1}(g_1,g_3) \cong \left\{ (q_{12},q_{23})\in C_\alpha\times C_\beta \ \middle|\ q_{12}q_{23}=r \right\}\,,
\end{equation}
which is generically a circle. The endpoints of $I_{\alpha,\beta}$ are treated separately, since the fibre type may change there.

With this notation the fusion operation \eqref{eq:fusion-bibranes} is schematically summarised by 
\begin{equation}
\raisebox{3pt}{\RawFusionDiagram} \;\;\; \raisebox{3pt}{\FusionArrow} \;\;\; \raisebox{3pt}{\FusedFusionDiagram}\,.
\end{equation}

Let
\begin{equation}
m:C_\alpha\times C_\beta\longrightarrow SU(2)\,, \qquad m(q_{12},q_{23})=q_{12}q_{23}\,,
\end{equation}
be the group multiplication map, and denote its restriction to $m^{-1}(C_\gamma)$ by $m_\gamma$. On this locus, the multiplicative identity for the conjugacy-class two-forms reads 
\begin{equation}\label{eq:conjugacy-class-fusion-form}
m_\gamma^*\omega_\gamma = {\rm pr}_1^*\omega_\alpha + {\rm pr}_2^*\omega_\beta -\frac{k}{2}\, \left\langle q_{12}^{-1}dq_{12} \wedge dq_{23}q_{23}^{-1} \right\rangle\,.
\end{equation}
Since
$g_1=q_{12}g_2$, $g_2=q_{23}g_3$, and $g_1=rg_3$, one finds
\begin{equation}\label{eq:mixed-term-identity-bibrane}
\left\langle r^{-1}dr\wedge dg_3g_3^{-1} \right\rangle = \left\langle q_{12}^{-1}dq_{12}\wedge dg_2g_2^{-1} \right\rangle + \left\langle q_{23}^{-1}dq_{23}\wedge dg_3g_3^{-1} \right\rangle - \left\langle q_{12}^{-1}dq_{12}\wedge dq_{23}q_{23}^{-1} \right\rangle\,.
\end{equation}
 On the locus \eqref{eq:clean-locus-bibrane}, the raw two-form is
\begin{equation}
F_{\alpha\beta}^{\rm raw} = \omega_\alpha(q_{12})+\omega_\beta(q_{23})- \frac{k}{2} \left\langle q_{12}^{-1}dq_{12}\wedge dg_2g_2^{-1} \right\rangle - \frac{k}{2} \left\langle q_{23}^{-1}dq_{23}\wedge dg_3g_3^{-1} \right\rangle\,.
\end{equation}
The standard two-form on $B_\gamma$, written in terms of $r=q_{12}q_{23}=g_1g_3^{-1}$ and $g_3$, is
\begin{equation}\label{eq:F-raw-gamma}
F_\gamma = \omega_\gamma(r) - \frac{k}{2} \left\langle r^{-1}dr\wedge dg_3g_3^{-1} \right\rangle .
\end{equation}
Using \eqref{eq:conjugacy-class-fusion-form} and \eqref{eq:mixed-term-identity-bibrane}, its pullback becomes
\begin{align}
\pi_{13}^*F_\gamma &=
\omega_\alpha+\omega_\beta - \frac{k}{2} \left\langle q_{12}^{-1}dq_{12}\wedge dq_{23}q_{23}^{-1} \right\rangle \\
 &\qquad- \frac{k}{2} \left[ \left\langle q_{12}^{-1}dq_{12}\wedge dg_2g_2^{-1} \right\rangle + \left\langle q_{23}^{-1}dq_{23}\wedge dg_3g_3^{-1} \right\rangle - \left\langle q_{12}^{-1}dq_{12}\wedge dq_{23}q_{23}^{-1} \right\rangle \right] \nonumber\,,
\end{align}
The two  $q_{12}$--$q_{23}$-terms cancel, leaving precisely $F_{\alpha\beta}^{\rm raw}$. The preceding identities therefore give
\begin{equation}\label{eq:bibrane-curvature-descends}
\left. F_{\alpha\beta}^{\rm raw} \right|_{\pi_{13}^{-1}(B_\gamma)} = \pi_{13}^*F_\gamma\,.
\end{equation}
Thus the raw two-form is basic for the restricted projection and descends to the standard bi-brane two-form $F_\gamma$. Since $(B_\gamma,F_\gamma)$ is the standard symmetry-preserving topological WZW bi-brane, every regular descended branch is topological.

%%%%%%%%%%%
\paragraph{Resulting fusion.}
At the level of support, the composition leads to a support going over the full interval of biconjugacy classes,
\begin{equation}
\pi_{13}\bigl(Y_{\alpha\beta}^{\rm raw}\bigr) = \bigcup_{\gamma\in I_{\alpha,\beta}}B_\gamma\,.
\end{equation}
This agrees with the classical product of conjugacy classes discussed in \cite{Fuchs:2007fw}, where the resulting interval was related to the bounds appearing in the affine fusion rules (see comments below). Here we obtain the same support from the fibre-product projection. We now evaluate the global holonomy condition on the factorisation fibres.

For $0<\alpha,\beta<\pi$ and $\gamma\in\mathrm{int}\,I_{\alpha,\beta}$, the factorisation fibre is a circle and the restricted projection satisfies the regularity assumptions used above. The curvature calculation therefore gives the regular local candidate branches
\begin{equation} 
\left\{ (B_\gamma,F_\gamma)\ \middle|\gamma\in\mathrm{int}I_{\alpha,\beta} \right\}\,.
\end{equation}
At the endpoints of $I_{\alpha,\beta}$ the circle fibre degenerates, so these loci are not covered by the regular reduction.

Suppose now that the input defects carry compatible global connection data. For each $\gamma\in\mathrm{int}I_{\alpha,\beta}$, the composite connection is flat along the factorisation circle, but it descends only when its holonomy around that circle is trivial. We therefore define
\begin{equation}
\mathcal A_{\alpha,\beta}^{\rm reg} :=\left\{\gamma\in\mathrm{int}\,I_{\alpha,\beta}\ \middle|\ \mathrm{Hol}_{\nabla_{\alpha\beta}^{\rm raw}}(\ell_{\rm fib})=1\right\}\,,
\end{equation}
where $\ell_{\rm fib}$ generates the factorisation circle over $B_\gamma$. In the notation of Section \ref{sec:classical_fusion}, the regular connection-level fusion is then
\begin{equation}\label{eq:bibrane-fusion-clean-strata}
\left. (B_\alpha,F_\alpha)\star(B_\beta,F_\beta)\right|_{\rm reg}=\bigsqcup_{\gamma\in\mathcal A_{\alpha,\beta}^{\rm reg}}(B_\gamma,F_\gamma)\,,
\end{equation}
with the descended connection data understood on every branch.

Let us consider two special cases from the degenerate bi-branes defined in \eqref{eq:special-case-bibranes}. For $\alpha=0$, the conjugacy class $C_0$ consists only of the identity element, so $q_{12}=\mathbf 1$, the factorisation fibre is trivial, and
\begin{equation}
C_0C_\beta=C_\beta\,, \qquad (B_0,F_0)\star(B_\beta,F_\beta) = (B_\beta,F_\beta)\,.
\end{equation}
Likewise, for $\alpha=\pi$, one has $C_\pi=\{-\mathbf 1\}$, hence $q_{12}=-\mathbf 1$, the fibre is again trivial, and
\begin{equation}
C_\pi C_\beta=C_{\pi-\beta}\,, \qquad (B_\pi,F_\pi)\star(B_\beta,F_\beta) = (B_{\pi-\beta},F_{\pi-\beta})\,.
\end{equation}
In both cases there are no vertical cycles and hence no additional fibre-holonomy obstruction to verify. Thus for this classical fusion, $B_0$ acts as the identity defect, while $B_\pi$ acts by multiplication with the non-trivial central element of $SU(2)$.

%%%%%%%%%%%%%%%%%
\paragraph{Relation to quantum fusion.}
After prequantisation and quantisation, the symmetry-preserving bi-branes correspond to the Verlinde defects of the diagonal WZW model. At level $k$ these are labelled by integrable weights $a,b,c\in\{0,\ldots,k\}$, with the usual quantum shift in the relation between the weight and the conjugacy-class position. Quantisation would then map fusion to multiplication in the Verlinde algebra, for which the fusion coefficients are
\begin{equation}
N_{ab}{}^c\neq0 \quad\Longleftrightarrow\quad |a-b|\leq c\leq\min(a+b,2k-a-b)\,, \quad a+b+c\in2\mathbb Z\,,
\end{equation}
see, for example, \cite{meinrenken2006lectures}. The local calculation above thus reproduces the classical support interval. This interval should then collapse into  the discrete set of labels upon imposing prequantisation conditions. We will further comment on the quantisation of this data and its challenges in the conclusions.

%%%%%%%%%%%%%%%%%%%%%%%%%%%%%%%%%%%%%%%%%%%%%%%%
\subsubsection{Corresponding local defect one-form}

We now describe the local one-form entering the defect action of the symmetry-preserving $SU(2)$ bi-brane.  Write a point of the bi-brane as before by $g_1=qg$, $g_2=g$ and $q\in C_\alpha$. In these coordinates, the intrinsic two-form \eqref{eq:Falpha-standard} becomes
\begin{equation}\label{eq:Falpha-qg-coordinates}
F_\alpha = \mathrm{pr}_{C_\alpha}^*\omega_\alpha -\frac{k}{2}\, \left\langle q^{-1}dq\wedge dg\,g^{-1}\right\rangle .
\end{equation}
For completeness, parameterise again the conjugacy class as in \eqref{eq:galpha-n} by
\begin{equation}
q=\cos\alpha\,\mathbf 1 +i\sin\alpha\,\hat n\cdot\sigma\,, \quad \hat n = (\sin\vartheta\cos\varphi,  \sin\vartheta\sin\varphi,  \cos\vartheta)\,,
\end{equation}
where $0\leq\vartheta\leq\pi$ and $\varphi\sim\varphi+2\pi$. The canonical conjugacy-class two-form  \cite{Alekseev:1998mc,Fuchs:2007fw} in the present normalisation takes the form
\begin{equation}
\omega_\alpha = -\frac{k}{4\pi}\sin(2\alpha)\, \sin\vartheta\,d\vartheta\wedge d\varphi\,.
\end{equation}
On a sufficiently small product patch on which the local potentials $b(q)$, $b(g)$ and $b(qg)$ are defined, the Polyakov-Wiegmann identity implies the existence of a one-form $\lambda$ satisfying
\begin{equation}
b(qg)-b(q)-b(g) +\frac{k}{2} \left\langle q^{-1}dq\wedge dg\,g^{-1} \right\rangle = d\lambda(q,g)\,,
\end{equation}
where the pullbacks to $C_\alpha\times SU(2)$ are understood.
Moreover, on a contractible patch of $C_\alpha$, choose a one-form $a_\alpha$ satisfying $da_\alpha = \omega_\alpha-\iota_\alpha^*b$, with $\iota_\alpha:C_\alpha\hookrightarrow SU(2)$. The local defect connection corresponding to \eqref{eq:Falpha-qg-coordinates} is then
\begin{equation}\label{eq:bibrane-local-one-form}
A_\alpha = \mathrm{pr}_{C_\alpha}^*a_\alpha-\lambda(q,g)\,.
\end{equation}
Indeed, eqs \eqref{eq:Falpha-qg-coordinates}--\eqref{eq:bibrane-local-one-form} imply $F_\alpha=p_1^*b-p_2^*b+dA_\alpha$, as required. The corresponding local defect contribution, as defined in the total action \eqref{eq:defect-action}, is therefore
\begin{equation}
S_{\mathrm{def}}^{(\alpha)} = \int_{\ell} A_\alpha = \int_{\ell} \left( \mathrm{pr}_{C_\alpha}^*a_\alpha-\lambda(q,g) \right)\,,
\end{equation}
where $\ell$ denotes the oriented defect line  and in the first equality the pullback of the one-form to $\ell$ is left understood.

For example, in a local $b$-field gauge for which $\iota_\alpha^*b=0$ on the northern patch of $C_\alpha\simeq S^2$, one may take
\begin{equation}
a_\alpha^{(N)}=-\frac{k}{4\pi}\sin(2\alpha)(1-\cos\vartheta)\,d\varphi\,.
\end{equation}
and hence in that local the defect action reads
\begin{equation}
S_{\mathrm{def}}^{(\alpha,N)} = \int_{\ell} \left[ -\frac{k}{4\pi}\sin(2\alpha) (1-\cos\vartheta)\,d\varphi -\lambda(q,g) \right]\,.
\end{equation}
Other choices of the local bulk potential $b$ change $A_\alpha$ by the corresponding defect gauge transformation.

%%%%%%%%%%%%%%%%%%%%%%%%%%%%%%%%%%%%%%%%%%%%%%%%%%%%%%
%%%%%%%%%%%%%%%%%%%%%%%%%%%%%%%%%%%%%%%%%%%%%%%%%%%%%%
\subsection{Hopf-fibre T-duality defect} \label{subsec:Hopf_T_duality}

In this subsection we construct the Hopf-fibre T-defect interface and computes its composition with the reverse defect. We first use the target-space topologicality conditions of Section \ref{sec:review} to determine the intrinsic defect two-form from the two bulk geometries. We then apply the fusion prescription of Section \ref{sec:classical_fusion}; the resulting branches reproduce the finite family of fibre-shift defects expected fusion result.

Let $X=S^3$ be equipped with its Hopf fibration over $S^2$, with fibre coordinate $\phi$. We consider the degree-$p$ Hopf-fibre T-duality defect, with $p\in\mathbb Z_{>0}$, from $X$ to $\widetilde X=S^2\times S^1$ carrying $p$ units of $H$-flux. For $p=1$ this is the ordinary Abelian T-duality interface. For $p>1$ it is obtained by combining $\mathbb Z_p$ half-space gauging with T-duality \cite{Arias-Tamargo:2025xdd}. Its composition with the reverse interface will be shown below to be non-invertible. Target-space descriptions of Poisson-Lie T-duality defect, which reduces consistently to this Abelian T-duality defect, were derived in \cite{Demulder:2022nlz}.

We begin by specifying the details of the local geometry of the Hopf-fibre T-dual pair. Let $X=S^3$ be equipped with its Hopf fibration over $S^2$, with fibre coordinate $\phi$. We consider the degree-$p$ Hopf-fibre T-duality defect, with $p\in\mathbb Z_{>0}$, from $X$ to $\widetilde X=S^2\times S^1$ carrying $p$ units of $H$-flux. For $p=1$ this is the ordinary Abelian T-duality defect. For $p>1$ it is the degree-$p$ defect obtained by combining $\mathbb Z_p$ half-space gauging with T-duality \cite{Arias-Tamargo:2025xdd}. As we show below its composition with the reverse defect decomposes into several defects and is therefore non-invertible. Target-space descriptions of Poisson-Lie T-duality defects were also developed in \cite{Demulder:2022nlz}.

We begin with the local geometry of the Hopf-fibre T-dual pair. On the left we consider again the Hopf-fibration metric on $S^3$, 
\begin{equation}\label{eq:left_Hopf_metric}
ds_L^2 = \frac{R_L^2}{8} \left( (2\,d\phi-\cos\theta\,d\psi)^2+d\theta^2+\sin^2\theta\,d\psi^2 \right),
\end{equation}
with Hopf coordinates $\theta\in[0,\pi]$, $\psi\sim\psi+2\pi$, $\phi\sim\phi+2\pi$. It is convenient to introduce the fibre one-form $\xi_L:=2\,d\phi-\cos\theta\,d\psi$. On the right we take the corresponding degree-$p$ dual background, which becomes the trivially fibered $S^2\times S^1$, with
metric
\begin{equation}\label{eq:right_Hopf_metric}
ds_R^2 = \frac{R_L^2}{8}\Bigl(d\theta^2+\sin^2\theta\,d\psi^2\Bigr) + \frac{R_R^2}{2}\,d\widetilde\phi^2\,,
\end{equation}
where $\widetilde\phi\sim \widetilde\phi+2\pi$, and  $H$-fluxes on each side are
\begin{equation}\label{eq:dual_H_flux}
 H_L=0\,, \qquad H_R = \frac{p}{4\pi}\sin\theta\,d\theta\wedge d\psi\wedge d\widetilde\phi \,,
\end{equation}
where $p$ denotes the units of Kalb-Ramond flux. The angular coordinates $\phi$ and $\widetilde\phi$ have period $2\pi$. Consequently, the physical fibre radii in the normalisation of \eqref{eq:left_Hopf_metric} and \eqref{eq:right_Hopf_metric} are $r_L=\frac{R_L}{\sqrt{2}}$ and  $r_R=\frac{R_R}{\sqrt{2}}$. 

The natural defect worldvolume is the correspondence space
\begin{equation}\label{eq:Hopf_correspondence_space}
Y_T=X_L\times_{S^2}X_R\,, \quad \theta_L=\theta_R=\theta\,, \quad \psi_L=\psi_R=\psi\,,
\end{equation}
with local coordinates $(\theta,\psi,\phi,\widetilde\phi)$. 

We determine the defect two-form by imposing the target-space topologicality conditions reviewed in Section \ref{sec:review}. Within the invariant local ansatz analysed in Appendix \ref{app:Hopf_topologicality_details}, the curvature identity fixes the coefficient of the mixed fibre term, the kernel condition eliminates the remaining invariant base two-form, and the quotient condition fixes the fibre radii. The result is
\begin{equation}\label{eq:FT_Hopf}
F_T = -\frac{p}{2\pi} \left( d\phi-\frac12\cos\theta\,d\psi \right)\wedge d\widetilde\phi
\end{equation}
together with
\begin{equation}\label{eq:Hopf_radius_relation}
r_Lr_R=\frac{p}{2\pi}\,,  
\end{equation}
or equivalently $R_LR_R=\frac{p}{\pi}$ with $r_L=\frac{R_L}{\sqrt2},\,r_R=\frac{R_R}{\sqrt2}$. The two-form \eqref{eq:FT_Hopf} is the degree-$p$ Poincar\'e curvature\footnote{The Poincar\'e line bundle is the standard line bundle over the product of a circle and its dual circle. Its connection provides the canonical coupling between the two angular coordinates. With both coordinates of period $2\pi$, its curvature may be written, up to orientation, as
\begin{equation}
	F_{\mathcal P}=\frac{1}{2\pi}\,d\phi\wedge d\widetilde\phi\,,\quad\frac{1}{2\pi}\int_{S^1_\phi\times S^1_{\widetilde\phi}}F_{\mathcal P}=1\,.
\end{equation}
As we will see, in the fusion calculation for the Hopf T-duality defects, its holonomy around the intermediate circle produces the $\mathbb Z_p$ branch condition.} on the fibre product, with the sign fixed by our orientation convention. It satisfies
\begin{equation}
dF_T = -\frac{p}{4\pi} \sin\theta\,d\theta\wedge d\psi\wedge d\widetilde\phi = -p_R^*H_R\,.
\end{equation}

Thus the support and intrinsic two-form of the interface are determined, within the invariant ansatz, by the two bulk geometries and the topological defect conditions. On each fibre torus, the restriction of $F_T$ is, up to the orientation sign, the curvature of the $p$th power of the Poincar\'e line bundle. Globally, $F_T$ is the corresponding twisted intrinsic two-form on the correspondence space, as in topological T-duality \cite{Bouwknegt:2003vb}.

%%%%%%%%%%%%%%%%%%%%%%%%%%%%%%%%%%%%%%%%%%%%%%%%
\subsubsection{Classical fusion of the Hopf-fibre T-defect} \label{subsec:classical_fusion_Hopf}

We now compose the degree-$p$ Hopf interface with its orientation reversed. The corresponding decomposition into fibre-shift defects is known for compact bosons \cite{Fuchs:2007tx} and was derived from the fused defect action in \cite{Niro:2022ctq}. Here we derive its nonlinear sigma-model realisation directly from the target-space descent prescription of Section \ref{sec:classical_fusion}.
 If $\tau$ exchanges the two target-space factors, the reversed interface carries $F_T^\vee=-\tau^*F_T$, and we consider
\begin{equation}\label{eq:fusion-Hopf-defect}
(Y_T,F_T)\star(Y_T^\vee,F_T^\vee)\,.
\end{equation}
Concretely, the first interface runs from $X_L$ to $\widetilde X_I$, and its reverse from $\widetilde X_I$ to $X_R$. Thus here $X_L$ and $X_R$ are copies of the fluxless Hopf fibred three-sphere, while $\widetilde X_I=S^2\times S^1_{\widetilde\phi}$ carries $p$ units of $H$-flux.

The raw fibre product over the intermediate target is
\begin{equation}
Y_{\mathrm{raw}} = Y_{L\widetilde I}\times_{\widetilde X_I}Y_{\widetilde I R} = \Bigl\{ \theta_L=\theta_I=\theta_R,\ \psi_L=\psi_I=\psi_R \Bigr\} \subset X_L\times \widetilde X_I\times X_R\,,
\end{equation}
with local coordinates $(\theta,\psi,\phi_L,\widetilde\phi_I,\phi_R)$. For the reversed
defect one has
\begin{equation}\label{eq:FT_dual_rev}
F_T^{\vee} = +\frac{p}{2\pi} \Bigl(d\phi_R-\frac12\cos\theta\,d\psi\Bigr)\wedge d\widetilde\phi_I. 
\end{equation}
Hence one can summarise the fusion operation \eqref{eq:fusion-Hopf-defect} schematically by  
\begin{equation}
\raisebox{3pt}{\RawHopfFusionDiagram}
\;\;\,
\raisebox{1pt}{\FusionArrow} \hspace{3pt}
\raisebox{3pt}{\FusedHopfFusionDiagram}\hspace{-1pt}.
\end{equation}

%%%%%%
\paragraph{Fused support and two-form.}

The raw intrinsic two-form
\begin{equation}
F_{\mathrm{raw}} = \mathrm{pr}_{L\widetilde I}^*F_T + \mathrm{pr}_{\widetilde I R}^*F_T^\vee
\end{equation}
is therefore
\begin{align}
F_{\mathrm{raw}} &= -\frac{p}{2\pi} \Bigl(d\phi_L-\frac12\cos\theta\,d\psi\Bigr)\wedge d\widetilde\phi_I + \frac{p}{2\pi} \Bigl(d\phi_R-\frac12\cos\theta\,d\psi\Bigr)\wedge d\widetilde\phi_I \notag\\
&= -\frac{p}{2\pi}\,d(\phi_L-\phi_R)\wedge d\widetilde\phi_I\,.\label{eq:Fraw_BF}
\end{align}
The Hopf connection terms cancel exactly, and in particular $dF_{\mathrm{raw}}=0$. On a local choice of angular lifts, a convenient primitive is
\begin{equation}\label{eq:Araw_fusion}
A_{\mathrm{raw}} = -\frac{p}{2\pi} (\phi_L-\phi_R)\,d\widetilde\phi_I\,, \qquad dA_{\mathrm{raw}}=F_{\mathrm{raw}}\,.
\end{equation}
We now project to the external targets,
\begin{equation}\label{eq:projection_LR}
\pi_{LR}:Y_{\mathrm{raw}}\to X_L\times X_R\,, \qquad \pi_{LR}(\theta,\psi,\phi_L,\widetilde\phi_I,\phi_R) = (\theta,\psi,\phi_L;\theta,\psi,\phi_R)\,.
\end{equation}
At the level of support,
\begin{equation}\label{eq:Yeff_set}
Y_{\mathrm{eff}} = \pi_{LR}(Y_{\mathrm{raw}}) = \Bigl\{ \theta_L=\theta_R,\ \psi_L=\psi_R \Bigr\} \subset X_L\times X_R\,.
\end{equation} 
But, as explained in Section \ref{sec:classical_fusion}, the projected support alone is not the fused defect: one must check the local curvature condition for $F_{\mathrm{raw}}$ and the global descent condition for the raw connection represented locally by $A_{\mathrm{raw}}$.

The fibres of $\pi_{LR}$ are $U(1)$-fibers parametrised by $\widetilde\phi_I$, so the vertical distribution is generated by $V=\partial_{\widetilde\phi_I}\in\ker(d\pi_{LR})$. The horizontality condition gives
\begin{equation}
\iota_{\partial_{\widetilde\phi_I}}F_{\mathrm{raw}} = -\frac{p}{2\pi}\, \iota_{\partial_{\widetilde\phi_I}} \bigl(d(\phi_L-\phi_R)\wedge d\widetilde\phi_I\bigr) = \frac{p}{2\pi}\,d(\phi_L-\phi_R)\,.
\end{equation}
Thus $F_{\mathrm{raw}}$ does not descend on all of $Y_{\mathrm{raw}}$. It becomes horizontal after restricting to a level set\footnote{Although $\phi_L$ and $\phi_R$ are only local Hopf coordinates, they undergo the same transition on the fibre product. Hence $\Delta\phi$ is a globally defined $S^1$-valued function. Local real lifts differ by $\Delta\phi\mapsto\Delta\phi+2\pi m$. Under this change,
\begin{equation}
A_{\mathrm{raw}}\longmapsto A_{\mathrm{raw}}-p\,m\,d\widetilde\phi_I\,,
\end{equation}
which is a connection gauge transformation because $p,m\in\mathbb Z$. The fibre holonomy is therefore independent of the chosen lift.}
\begin{equation}
\Delta\phi:=\phi_L-\phi_R=\mathrm{const}\,.
\end{equation}
The allowed constants are then determined by the global holonomy condition.

Horizontality ensures the descent of the curvature on each level set. Global descent of the connection additionally requires trivial holonomy around the intermediate circle. Restricting \eqref{eq:Araw_fusion} to a fibre $S^1_{\widetilde\phi_I}$ gives
\begin{equation}
A_{\mathrm{raw}}\big|_{S^1_{\widetilde\phi_I}} = -\frac{p}{2\pi}\,\Delta\phi\, d\widetilde\phi_I\,.
\end{equation}
Since $\widetilde\phi_I\sim \widetilde\phi_I+2\pi$, trivial fibre holonomy requires
\begin{equation}
\exp\left( i\oint_{S^1_{\widetilde\phi_I}}A_{\mathrm{raw}} \right) = \exp(-ip\,\Delta\phi) = 1\,.
\end{equation}
Thus we obtain the condition 
\begin{equation}
\Delta\phi\in-\frac{2\pi}{p}\mathbb Z \qquad\bmod 2\pi\,.
\end{equation}
The inequivalent values are therefore labelled by $\mathbb Z/p\mathbb Z\cong\mathbb Z_p$. For $\eta\in\mathbb Z_p$, choose any integer representative $\widehat\eta\in\mathbb Z$. The corresponding admissible branch is
\begin{equation}\label{eq:Y_eta}
Y_\eta=\left\{\theta_L=\theta_R,\quad\psi_L=\psi_R,\quad\phi_L=\phi_R-\frac{2\pi}{p}\widehat\eta\right\}\subset X_L\times X_R\,.
\end{equation}
This definition is independent of the representative, since $\widehat\eta\mapsto\widehat\eta+p$ shifts the right-hand side by $2\pi$. Each $Y_\eta$ is the graph of the fibre rotation
\begin{equation}\label{eq:fibre_shift_eta}
(\theta,\psi,\phi)\longmapsto \left(\theta,\psi,\phi+\frac{2\pi}{p}\widehat\eta\right)\,.
\end{equation}

On the admissible locus $\pi_{LR}^{-1}(Y_\eta)$, the difference $\phi_L-\phi_R$ is constant. Hence
\begin{equation}
F_{\mathrm{raw}}\big|_{\pi_{LR}^{-1}(Y_\eta)}=0\,, \quad A_{\mathrm{raw}}\big|_{\pi_{LR}^{-1}(Y_\eta)} =\widehat\eta\,d\widetilde\phi_I\,.
\end{equation}
Because $\widehat\eta\in\mathbb Z$, the gauge transformation $e^{-i\widehat\eta\widetilde\phi_I}$ is single-valued. The raw connection is therefore gauge-equivalent to the pullback of the trivial connection and descends to $Y_\eta$. 

%%%%%%
\paragraph{Resulting fusion.}

Taken altogether, the composition of the Hopf T-duality defect with its reverse therefore decomposes classically as
\begin{equation}\label{eq:Hopf_defect_fusion}
(Y_T,F_T)\star(Y_T^\vee,F_T^\vee) = \bigsqcup_{\eta\in\mathbb Z_p}(Y_\eta\,,0)\,.
\end{equation}
In particular, the branching is already visible before quantisation: it is the classical consequence of the descent of the raw connection data through the compact intermediate fibre. Each $Y_\eta$ is the graph of a fibre rotation preserving the metric and three-form flux. Together with $F_\eta=0$, this verifies that $(Y_\eta,0)$ is a topological defect. We will come back to this family of defects in the next subsection.

As mentioned above, this fusion rule is known. In the present calculation, however, the origin of the discrete branches is made explicit by connection-level descent:  curvature horizontality fixes $\Delta\phi$ to be constant, while fibre holonomy restricts it to
\begin{equation}
\Delta\phi = -\frac{2\pi}{p}\eta \;\bmod 2\pi\,, \quad \eta\in\mathbb Z_p\,.
\end{equation}
As the following examples illustrate, the same prescription systematically determines the admissible fusion branches together with their descended two-form and connection data.

%%%%%%%%%%%%%%%%%%%%%%%%%%%%%%%%%%%%%%%%%%%%%%%%
\subsubsection{Corresponding local defect one-form} \label{sec:Hopf_one_form_action}

We orient the defect line such that $\gamma=\partial\Sigma_L=-\partial\Sigma_R$ and write, as before, $S_{\mathrm{def}}^{(T)}=\int_\gamma A_T$, see \eqref{eq:defect-action}. 
On a contractible target-space patch, the intrinsic two-form and the local bulk potentials are related by
\begin{equation}\label{eq:Hopf_a_b_relation}
F_T=p_L^*b_L-p_R^*b_R+dA_T, \quad db_L=H_L,\quad db_R=H_R\,.
\end{equation}
We use $p\in\mathbb Z_{>0}$ throughout. A convenient local gauge is
\begin{equation}\label{eq:bR_Hopf_choice}
b_L=0,\qquad b_R=-\frac{p}{4\pi}\cos\theta\, d\psi\wedge d\widetilde\phi\,.
\end{equation}
Substituting \eqref{eq:FT_Hopf} into \eqref{eq:Hopf_a_b_relation} gives
\begin{equation}\label{eq:da_Hopf}
dA_T=-\frac{p}{2\pi}\, d\phi\wedge d\widetilde\phi\,.
\end{equation}
We may therefore choose $A_T=-\frac{p}{2\pi}\phi\,d\widetilde\phi$ and hence
\begin{equation}\label{eq:Sdef_Hopf}
S_{\mathrm{def}}^{(T)} =-\frac{p}{2\pi}\int_\gamma \phi\,d\widetilde\phi\,.
\end{equation}
The choice $b_L=0$ is global on the fluxless left-hand background, whereas $b_R$ and $A_T$ are only locally defined. 
On an overlap, if $b_R^{(i)}-b_R^{(j)}=d\Lambda_{R,ij}$, for some one-form $\Lambda_{R,ij}$, then
\begin{equation}
A_T^{(i)}-A_T^{(j)} = p_R^*\Lambda_{R,ij}+d\chi_{ij}\,,
\end{equation}
so that \eqref{eq:Hopf_a_b_relation} is indeed independent of the local choices. In particular, changing a local lift by $\phi\mapsto\phi+2\pi m$ shifts $A_T$ by $-pm\,d\widetilde\phi$. It is an allowed connection gauge transformation because $p\in\mathbb Z$. On each local fibre torus, \eqref{eq:da_Hopf} is the degree-$p$ Poincar\'e curvature, up to orientation.

%%%%%%%%%%%%%%%%%%%%%%%%%%%%%%%%%%%%%%%%%%%%%%%%
\subsubsection{TsT correspondences from fusion}\label{sec:TsT_fusion}

TsT transformations are known in holography as the string-theory realisation of the real $\beta$-deformation of $\mathcal N=4$ super Yang-Mills theory \cite{Lunin:2005jy,Frolov:2005dj}. Such a transformation can be performed on a sigma model with two commuting circle isometries by T-dualising along one circle, shifting the second coordinate by the dual coordinate, and T-dualising back. In the examples relevant here, this construction generates continuous families of classically integrable sigma models; the associated Lax connection and its relation to twisted boundary conditions were studied in \cite{Frolov:2005dj,Alday:2005ww}. More generally, commuting TsT transformations are equivalent to Abelian homogeneous Yang-Baxter deformations and may be understood as a distinguished class of $O(d,d)$ $\beta$-transformations \cite{Osten:2016dvf,Orlando:2019his}. Here we reinterpret the three elementary steps of TsT as a composition of classical defects and determine both the resulting target-space correspondence and the conditions for its global connection data to descend.

To implement this construction in the Hopf example, we enlarge the fluxless target by a spectator circle and consider
\begin{equation}\label{eq:tst-targets}
X=S^3_{\rm Hopf}\times S^1_\chi\,,\qquad \widetilde X =S^2\times S^1_{\widetilde\phi}\times S^1_\chi\,.
\end{equation}
All angular coordinates have period $2\pi$. Introduce the global Hopf connection
\begin{equation}
\vartheta= d\phi-\frac12\cos\theta\,d\psi\,, \quad d\vartheta=\frac12\sin\theta\,d\theta\wedge d\psi\,,
\end{equation}
and write the original background as
\begin{equation}
g_0=g_{S^2}+r^2\vartheta^2+\rho^2d\chi^2\,,\quad b_0=0\,,\quad H_0=0\,.
\end{equation}
Here $r$ and $\rho$ are the physical radii of the Hopf and spectator circles. We set
\begin{equation}
\lambda:=\frac{p}{2\pi}\,,\quad r\widetilde r=\lambda\,,
\end{equation}
where $\widetilde r$ is the radius of the T-dual circle.

After the first T-duality, one implements the intermediate shift locally by introducing a real parameter $\zeta$ and setting
\begin{equation}\label{eq:genuine_TsT_shear}
\widetilde\phi_M=\widetilde\phi_I\,,\quad \chi_M=\chi_I+\lambda\zeta\,\widetilde\phi_I\,,\quad \zeta\in\mathbb R\,.
\end{equation}
The first T-dual background is
\begin{equation}
\widetilde g_I=g_{S^2}+\widetilde r^2d\widetilde\phi_I^2+\rho^2d\chi_I^2\,,\qquad\widetilde b_I=\lambda\mathcal A\wedge d\widetilde\phi_I\,,
\end{equation}
where
\begin{equation}
\mathcal A= -\frac12\cos\theta\,d\psi\,, \quad \vartheta=d\phi+\mathcal A\,.
\end{equation}
In the sheared coordinates, the intermediate background hence becomes
\begin{equation}
\widetilde g_M =g_{S^2}+\widetilde r^2d\widetilde\phi_M^2+\rho^2\left(d\chi_M-\lambda\zeta\,d\widetilde\phi_M\right)^2\,,\quad \widetilde b_M = \lambda\mathcal A\wedge d\widetilde\phi_M\,.
\end{equation}

Let $X_\zeta$ denote the background obtained by T-dualising$\widetilde X_M$ along $\widetilde\phi_M$, and define $\Delta_\zeta:=1+\zeta^2r^2\rho^2$. The components involving the dualised coordinate are
\begin{equation}
(\widetilde g_M)_{\widetilde\phi\widetilde\phi}=\widetilde r^2+\lambda^2\zeta^2\rho^2=\widetilde r^2\Delta_\zeta\,,\quad(\widetilde g_M)_{\widetilde\phi\chi}=-\lambda\zeta\rho^2\,.
\end{equation}
Since the angular coordinates have period $2\pi$ and $r\widetilde r=\lambda$, the Buscher rule in our normalisation gives
\begin{equation}
(g_\zeta)_{\phi\phi}=\frac{\lambda^2}
     {(\widetilde g_M)_{\widetilde\phi\widetilde\phi}}=\frac{r^2}{\Delta_\zeta}\,.
\end{equation}
Applying the remaining Buscher rules \cite{Buscher:1987qj} gives the background fields\footnote{For completeness; in the same normalisation, the two T-dualities shift the dilaton according to
\begin{equation}
\widetilde\Phi_I=\Phi_0-\frac12\log\left(\frac{r^2}{\lambda}\right)\,, \quad \Phi_\zeta=\widetilde\Phi_I-\frac12\log\left( \frac{\widetilde r^2\Delta_\zeta}{\lambda}\right)\,.
\end{equation}
Using $r\widetilde r=\lambda$, the net shift is $\Phi_\zeta = \Phi_0-\frac12\log\Delta_\zeta$ \cite{Buscher:1987qj,Tseytlin:1991wr,Schwarz:1992te}, however the dilaton shift is not needed for the classical defect conditions we consider here.} 
\begin{align}
g_\zeta &=g_{S^2}+\frac{r^2}{\Delta_\zeta}\,\vartheta_R^2+\frac{\rho^2}{\Delta_\zeta}\,d\chi_R^2\,,\label{eq:TsT_metric}\\
b_\zeta &=\frac{\zeta r^2\rho^2}{\Delta_\zeta}\,\vartheta_R\wedge d\chi_R\,,\label{eq:TsT_b_field}\\
H_\zeta &=db_\zeta=\frac{\zeta r^2\rho^2}{2\Delta_\zeta}\sin\theta\,d\theta\wedge d\psi\wedge d\chi_R\,,\label{eq:TsT_H_flux}
\end{align}
where $\vartheta_R=d\phi_R+\mathcal A$.  We choose the orientation of the dual coordinate consistently with $\widetilde b_I=\lambda\mathcal A\wedge d\widetilde\phi_I$; reversing this convention is equivalent to replacing $\zeta$ by $-\zeta$.

Hence for any non-zer0 $\zeta\neq0$, we consider the local threefold composition
\begin{equation}\label{eq:TsT-threefold-composition}
(Y_T,F_T)\star_{\mathrm{loc}}(Y_{S_\zeta},0)\star_{\mathrm{loc}}(Y_{T_\zeta}^{\vee},F_{T_\zeta}^{\vee})\,.
\end{equation}
Here $\star_{\mathrm{loc}}$ denotes reduction of the support and intrinsic two-form before global connection descent. It becomes the full fusion $\star$ only when the connection data also descend. As shown below, the resulting data satisfy the pointwise topological defect conditions for every $\zeta\neq0$, while for generic $\zeta$ the rank-one connection does not descend globally.

Schematically, the composition is
\begin{equation}\label{eq:TsT-fusion-diagram}
\RawTsTFusionDiagram \;\xrightarrow{\;\hspace{15pt}\;}\;\FusedTsTFusionDiagram\,,
\end{equation}
where the arrow denotes local reduction at this stage. The raw fibre product represented on the left-hand side is subject to
\begin{equation}\label{eq:TsT-raw-identifications}
\begin{gathered}
(\theta,\psi)_L=(\theta,\psi)_I=(\theta,\psi)_M=(\theta,\psi)_R\,,\\
\widetilde\phi_I=\widetilde\phi_M=:\widetilde\phi\,,\quad \chi_R-\chi_L= \lambda\zeta\,\widetilde\phi\,.
\end{gathered}
\end{equation}

%%%%%%
\paragraph{Fused support and two-form.}

Eliminating the intermediate coordinate $\widetilde\phi$ from \eqref{eq:TsT-raw-identifications}, the local projection to the external targets has, for $\zeta\neq0$, the image
\begin{equation}\label{eq:TsT_projected_support}
Y_\zeta^{\rm TsT}= \left\{\theta_L=\theta_R,\;\psi_L=\psi_R\right\}\subset X_L\times X_{\zeta,R}\,.
\end{equation}
There is no constraint to be imposed on either $\phi_L-\phi_R$ or $\chi_R-\chi_L$. In the local conventions fixed above, the two T-duality connections are
\begin{equation}
A_T=-\lambda\phi_L\,d\widetilde\phi\,,\quad A_{T_\zeta}^{\vee}= +\lambda\phi_R\,d\widetilde\phi\,.
\end{equation}
The shear graph carries the trivial local connection, so the raw connection is $A_{\rm TsT}^{\rm raw} = -\lambda(\phi_L-\phi_R)\,d\widetilde\phi$. Using $d(\chi_R-\chi_L)= \lambda\zeta\,d\widetilde\phi$, we obtain the local one-form
\begin{equation}\label{eq:genuine_TsT_connection}
A_\zeta=-\frac{1}{\zeta}(\phi_L-\phi_R)\,d(\chi_R-\chi_L)\,.
\end{equation}
The corresponding intrinsic two-form on $Y_\zeta^{\rm TsT}$ is
\begin{align}\label{eq:TsT_intrinsic_form}
F_\zeta&=p_L^*b_0-p_R^*b_\zeta+dA_\zeta =-\frac{\zeta r^2\rho^2}{\Delta_\zeta}\,\vartheta_R\wedge d\chi_R-\frac{1}{\zeta}\,d(\phi_L-\phi_R)\wedge d(\chi_R-\chi_L)\,.
\end{align}
It obeys $dF_\zeta=p_L^*H_0-p_R^*H_\zeta$, as required.

%%%%%%
\paragraph{Topological check.}

We next verify the classical topologicality conditions. Define the two-dimensional distribution
\begin{equation}
\mathcal K_\zeta=\ker\vartheta_L\cap\ker\vartheta_R\cap\ker d\chi_L\cap\ker d\chi_R\subset TY_\zeta^{\rm TsT}\,.
\end{equation}
A direct restriction of the neutral metric shows that $\mathcal K_\zeta\subseteq\bigl(TY_\zeta^{\rm TsT}\bigr)^\perp$. Both distributions have rank two, and hence
\begin{equation}
\mathcal K_\zeta=\bigl(TY_\zeta^{\rm TsT}\bigr)^\perp\,.
\end{equation}
Direct contraction of \eqref{eq:TsT_intrinsic_form} gives $\iota_VF_\zeta=0$ for every $V\in\mathcal K_\zeta$, so $F_\zeta$ descends to the quotient. The descended form displayed below is nondegenerate, and therefore
\begin{equation}
\ker F_\zeta=\mathcal K_\zeta= \bigl(TY_\zeta^{\rm TsT}\bigr)^\perp\,.
\end{equation}

On the quotient $TY_\zeta^{\rm TsT}/\mathcal K_\zeta$, use the coframe $\left(\vartheta_L,\,d\chi_L,\,\vartheta_R,\,d\chi_R\right)$. The descended neutral metric is
\begin{equation}
\widetilde G_\zeta= \mathrm{diag}\left( r^2,\,\rho^2,\,-\frac{r^2}{\Delta_\zeta},\,-\frac{\rho^2}{\Delta_\zeta}\right)\,,
\end{equation}
while the descended two-form is
\begin{equation}
\widetilde F_\zeta=-\frac{\zeta r^2\rho^2}{\Delta_\zeta}\,\vartheta_R\wedge d\chi_R-\frac{1}{\zeta}(\vartheta_L-\vartheta_R)\wedge(d\chi_R-d\chi_L)\,.
\end{equation}
Using $\Delta_\zeta=1+\zeta^2r^2\rho^2$, direct multiplication gives
\begin{equation}
\left(\widetilde G_\zeta^{-1}\widetilde F_\zeta\right)^2=\mathrm{id}\,.
\end{equation}
The pair $(Y_\zeta^{\rm TsT},F_\zeta)$ therefore satisfies the target-space topologicality criterion \eqref{eq:KS_geometric_conditions} for every $\zeta\in\mathbb R\setminus\{0\}$.

In order to promote this locally topological pair to a full connection-level defect in the sense of \eqref{eq:classical-fusion-prescription}, the local one-form $A_\zeta$ must define a globally well-defined rank-one connection. Its curvature must therefore have integral periods. On the two-torus parametrised by $\Delta\phi:=\phi_L-\phi_R$ and $\Delta\chi:=\chi_R-\chi_L$, one finds
\begin{equation}
\frac{1}{2\pi}\int_{S^1_{\Delta\phi}\times S^1_{\Delta\chi}}dA_\zeta=-\frac{2\pi}{\zeta}\,.
\end{equation}
A necessary integrality condition is consequently
\begin{equation}\label{eq:TsT_prequantization_condition}
\frac{2\pi}{\zeta}\in\mathbb Z\,.
\end{equation}

We now distinguish between the local composition $\star_{\mathrm{loc}}$ and the full connection-level fusion $\star$ defined in Section~\ref{sec:classical_fusion}. The local reduction is defined for arbitrary $\zeta\neq0$. For the threefold composition to define a full fusion of globally defined defects on fixed compact tori, the intermediate shear \eqref{eq:genuine_TsT_shear} must itself be globally defined. This requires
\begin{equation}\label{eq:TsT_shear_integrality}
\ell:= \lambda\zeta=\frac{p\zeta}{2\pi}\in\mathbb Z\,.
\end{equation}
For $\ell\neq0$, the projection of the raw correspondence is an $|\ell|$-sheeted covering of $Y_\zeta^{\rm TsT}$. Combining \eqref{eq:TsT_prequantization_condition} and \eqref{eq:TsT_shear_integrality} gives
\begin{equation}\label{eq:condition-divisibility}
\zeta=\frac{2\pi}{N}\,, \qquad N\in\mathbb Z\setminus\{0\}\,, \qquad N\mid p\,.
\end{equation}
On this locus, rank-one descent further requires compatible identifications of the raw line data over the sheets of the covering. Without such identifications, their natural pushforward has rank $|\ell|$. Thus $\star_{\mathrm{loc}}$ produces the globally specified pair $(Y_\zeta^{\rm TsT},F_\zeta)$ for arbitrary $\zeta\neq0$, whereas the full fusion $\star$ is defined on the discrete locus \eqref{eq:condition-divisibility}, after the required sheet identifications have been chosen.

%%%%%%
\paragraph{Resulting fusion.}

At the level of local support and intrinsic two-form data, the threefold composition gives
\begin{equation}\label{eq:genuine_TsT_local_fusion_law}
(Y_T,F_T) \star_{\mathrm{loc}} (Y_{S_\zeta},0)\star_{\mathrm{loc}} (Y_{T_\zeta}^{\vee},F_{T_\zeta}^{\vee}) = (Y_\zeta^{\rm TsT},F_\zeta)\,, \quad \zeta\neq0\,.
\end{equation}
The resulting pair satisfies the pointwise topological defect conditions, but for generic $\zeta$ it does not carry a globally descended rank-one connection on the compact targets.

On the locus \eqref{eq:condition-divisibility}, and provided compatible identifications of the connection data over the sheets of the covering are chosen, the full connection-level fusion is
\begin{gather}
\begin{aligned}\label{eq:genuine-TsT-global-fusion}
&(Y_T,F_T,\nabla_T)\star(Y_{S_\zeta},0,\nabla_{S_\zeta})\star (Y_{T_\zeta}^{\vee},   F_{T_\zeta}^{\vee},  \nabla_{T_\zeta}^{\vee})  = (Y_\zeta^{\rm TsT},F_\zeta, \nabla_\zeta)\,, \\
&\qquad \qquad \text{for}\quad  \zeta=\frac{2\pi}{N}\,, \quad N\in\mathbb Z\setminus\{0\}\,, \quad N\mid p\,.
\end{aligned}
\end{gather}
For $\zeta=0$, the intermediate shear is trivial and the composition reduces to the $\mathbb Z_p$-branched fusion of the Hopf interface with its reverse described in Section~\ref{subsec:Hopf_T_duality}.

%%%%%%%%%%%%%%%%%%%%%%%%%%%%%%%%%%%%%%%%%%%%%%%%
\subsubsection{Corresponding local defect one-form}

As before, the geometric data obtained from the fusion prescription determine a local defect one-form and the associated defect action. For the TsT correspondence, the local one-form is
\begin{equation}\label{eq:TsT_local_one_form}
A_\zeta = -\frac{1}{\zeta} (\phi_L-\phi_R)\, d(\chi_R-\chi_L)\,,
\end{equation}
and the corresponding local defect contribution is
\begin{equation}\label{eq:Sdef_genuine_TsT}
S_{\rm def}^{({\rm TsT},\zeta)} = -\frac{1}{\zeta} \int_\gamma(\phi_L-\phi_R)\, d(\chi_R-\chi_L)\,.
\end{equation}
Under a change of local lift
\begin{equation}
\phi_L-\phi_R \longmapsto \phi_L-\phi_R+2\pi n\,, \quad n\in\mathbb Z\,,
\end{equation}
the one-form transforms as
\begin{equation}
A_\zeta \longmapsto A_\zeta-\frac{2\pi n}{\zeta}\, d(\chi_R-\chi_L)\,.
\end{equation}
This is a globally admissible connection gauge transformation precisely when $2\pi/\zeta\in\mathbb Z$. Independently of this global integrality condition, the local relation $F_\zeta = -p_R^*b_\zeta+dA_\zeta$ holds for every $\zeta\neq0$.

%%%%%%%%%%%%%%%%%%%%%%%%%%%%%%%%%%%%%%%%%%%%%%%%
%%%%%%%%%%%%%%%%%%%%%%%%%%%%%%%%%%%%%%%%%%%%%%%%
\subsection{T-dual biconjugacy defects}\label{subsec:Tdual-biconjugacy-defects}

We now use the fusion prescription to compose a symmetry-preserving WZW bi-brane defect with a Hopf-fibre T-duality defect. The result yields a defect that separates on the worldsheet the action between the $SU(2)$ WZW model and its T-dual lens-space theory. In contrast to the preceding examples, eliminating the intermediate Hopf coordinate is generically two-valued and becomes degenerate on a special locus. This example therefore exemplifies, in a distinct way to the instances already explored, how non-invertibility can emerge in classical fusion composition. Indeed, here fusion will generically produces two local branches, whose structure also changes on a degenerate locus.

For the composition to be defined, the two defects must share the same intermediate sigma model. We take $X_L=X_I=X=SU(2)$ with its level-$k$ WZW background $(g,H)$. The Hopf bundle has degree one, while the fibre integral of the WZW flux is $k$. Topological T-duality exchanges these data, so the dual target is the lens space $\widetilde X_R=L(k,1)$ with dual background $(\widetilde g,\widetilde H)$ \cite{Bouwknegt:2003vb,Sarkissian:2008dq}. Denote the bundle projections by
\begin{equation}
\pi_X:X_I\longrightarrow S^2\,, \qquad \pi_{\widetilde X}:\widetilde X_R\longrightarrow S^2\,.
\end{equation}

On $X_L\times X_I$ we take the symmetry-preserving WZW bi-brane $(B_\alpha,F_\alpha)$ defined in \eqref{eq:Balpha-def}, while on $X_I\times\widetilde X_R$ we take the corresponding Hopf T-duality defect $(Y_T,F_T)$. Their intrinsic two-forms satisfy
\begin{align}
dF_\alpha&=p_L^*H-p_I^*H\,,\label{eq:Tdual-biconj-Falpha-twist}\\
dF_T&=p_I^*H-p_{\widetilde R}^*\widetilde H\,.\label{eq:Tdual-biconj-FT-twist}
\end{align}
The support of the T-duality defect is the correspondence space
\begin{equation}
Y_T =X_I\times_{S^2}\widetilde X_R=\left\{(g_I,\widetilde x_R)\ \middle|\pi_X(g_I)=\pi_{\widetilde X}(\widetilde x_R)\right\}\,.
\end{equation}
In local Hopf coordinates this becomes
\begin{equation}
Y_T= \left\{ (\theta_I,\psi_I,\phi_I;  \theta_R,\psi_R,\widetilde\phi_R) \ \middle|\ \theta_I=\theta_R,\quad\psi_I=\psi_R \right\}\,.
\end{equation}

We study the mixed composition
\begin{equation}
(B_\alpha,F_\alpha) \star_{\mathrm{loc}}(Y_T,F_T)\,,
\end{equation}
where $\star_{\mathrm{loc}}$ denotes reduction of the support and intrinsic two-form before global connection descent. Its set-up is summarised schematically by
\begin{equation}\label{eq:Tdual-biconj-setup-diagram}
\raisebox{3pt}{\TdualBiconjSetupDiagram}\,.
\end{equation}
This composition has a known quantum counterpart. Starting from symmetry-preserving WZW defect operators, \cite{Sarkissian:2008dq} constructed defects between the $SU(2)_k$ and lens-space theories by applying T-duality to one chiral sector. We compare their large-level support with the classical fusion result below.

%%%%%%
\paragraph{Raw support and its projection.}

The first step is to match the two defects through their common $SU(2)$ field. The resulting raw correspondence is
\begin{align}\label{eq:Tdual-biconj-raw-support}
Y_{\alpha T}^{\rm raw} &:= B_\alpha\times_{X_I}Y_T \subset X_L\times X_I\times\widetilde X_R \nonumber\\
&= \left\{ (g_L,g_I,\widetilde x_R) \ \middle|\ g_Lg_I^{-1}\in C_\alpha,\; \pi_X(g_I)=\pi_{\widetilde X}(\widetilde x_R) \right\}\,,
\end{align}
where $\pi_X:X\to S^2$ is the Hopf projection. Choose then local coordinates
\begin{equation}
g_L=g(\theta_L,\psi_L,\phi_L)\,, \qquad g_I=g(\theta_I,\psi_I,\phi_I)
\end{equation}
and $(\theta_R,\psi_R,\widetilde\phi_R)$ on $\widetilde X_R$, with $\theta_I=\theta_R$ and $\psi_I=\psi_R$. Using \eqref{eq:Balpha-Hopf-eq}, the biconjugacy constraint becomes
\begin{equation}\label{eq:Tdual-biconj-explicit-constraint}
\cos\alpha = A_{LR}\cos(\phi_L-\phi_I) + B_{LR}\sin(\phi_L-\phi_I)\,,
\end{equation}
where
\begin{align}
A_{LR} &= \cos\frac{\theta_L-\theta_R}{2} \cos\frac{\psi_L-\psi_R}{2}\,,\quad  B_{LR} = \cos\frac{\theta_L+\theta_R}{2} \sin\frac{\psi_L-\psi_R}{2}\,.
\end{align}

We now eliminate the intermediate coordinate by projecting to the
external targets:
\begin{equation}\label{eq:Tdual-biconj-projection}
\pi_{L\widetilde R}: Y_{\alpha T}^{\rm raw} \longrightarrow X_L\times\widetilde X_R\,, \quad (g_L,g_I,\widetilde x_R) \longmapsto (g_L,\widetilde x_R)\,.
\end{equation}
For later convenience, define $R_{LR}:=\sqrt{A_{LR}^2+B_{LR}^2}$. Equation \eqref{eq:Tdual-biconj-explicit-constraint} admits a solution for $\phi_I$ precisely when $\lvert\cos\alpha\rvert\leq R_{LR}$. The projected support is therefore
\begin{equation}\label{eq:Tdual-biconj-effective-support}
Y_{\alpha T}^{\rm eff} = \left\{ (g_L,\widetilde x_R)\in X_L\times\widetilde X_R \ \middle|\ \cos^2\alpha\leq A_{LR}^2+B_{LR}^2 \right\}\,.
\end{equation}
Equivalently, choosing a local lift
$\widehat g_R\in SU(2)$ of the lens-space coordinate, the existence of
an intermediate field $g_I$ can be written invariantly as
\begin{equation}\label{eq:Tdual-biconj-Sarkissian-support}
g_L^{-1}\widehat g_R\in C_\alpha U(1)\,,
\end{equation}
where $U(1)\subset SU(2)$ acts along the Hopf fibre. Thus \eqref{eq:Tdual-biconj-effective-support} is the local coordinate parametrisation of the large-level support found in \cite{Sarkissian:2008dq}.

Note that, in general the effective support \eqref{eq:Tdual-biconj-effective-support} is a subset with boundary, rather than a smooth embedded bi-brane worldvolume. We need to distinguish two cases. We will call the branches regular, when the defining constraint in \eqref{eq:Tdual-biconj-effective-support} yields two solutions.  Besides the regular locus there are also two kinds of singular behaviour. When $R_{LR}=|\cos\alpha|>0$, the two solutions coalesce and the projection develops a fold. When $R_{LR}=0$, solutions exist only for $\alpha=\pi/2$; in that case the constraint becomes independent of $\phi_I$ and the factorisation fibre is the full intermediate circle. We treat this exceptional reduction separately below.

%%%%%%
\paragraph{Regular branches.}

\begin{figure}[t]
\flushleft

\begin{minipage}[t]{0.62\textwidth}
\flushleft
\begin{tikzpicture}[
    x=1.15cm,
    y=1.35cm,
    axis/.style={
        black!35,
        line width=.45pt
    },
    curve/.style={
        black!70,
        line width=.85pt
    },
    level/.style={
        teal!90!black,
        line width=.75pt
    },
    guide/.style={
        black!28,
        densely dashed,
        line width=.4pt
    },
    label/.style={
        font=\scriptsize
    }
]

\def\deltaPlot{0.45}
\def\levelPlot{0.40}
\pgfmathsetmacro{\thetaPlot}{acos(\levelPlot)/180*pi}
\pgfmathsetmacro{\xMinus}{\deltaPlot-\thetaPlot}
\pgfmathsetmacro{\xPlus}{\deltaPlot+\thetaPlot}
\pgfmathsetmacro{\xLeft}{\deltaPlot-pi}
\pgfmathsetmacro{\xRight}{\deltaPlot+pi}

% Axes
\draw[axis,->]
    (\xLeft-.15,0) -- (\xRight+.28,0)
    node[below,label]
    {\hspace{-30pt}$x=\phi_L-\phi_I$};

\draw[axis,->]
    (\xLeft,-1.12) -- (\xLeft,1.20);

% Cosine constraint
\draw[curve,domain=\xLeft:\xRight,samples=120,smooth,variable=\x]
    plot
    (\x,{cos((\x-\deltaPlot) r)});

% Horizontal level
\draw[level]
    (\xLeft,\levelPlot) -- (\xRight,\levelPlot);

\node[label,anchor=south west, text=teal!90!black]
    at (\xLeft+.08,\levelPlot)
    {$\cos\alpha$};

% Two solutions
\draw[guide]
    (\xMinus,0) -- (\xMinus,\levelPlot);

\draw[guide]
    (\xPlus,0) -- (\xPlus,\levelPlot);

\fill[orange!85!black]
    (\xMinus,\levelPlot) circle (1.65pt);

\fill[blue!90!black]
    (\xPlus,\levelPlot) circle (1.65pt);

\node[label,anchor=north,align=center]
    at (\xMinus,-.05)
    {$x_-$};

\node[label,anchor=north,align=center]
    at (\xPlus,-.05)
    {$x_+$};

% Annotation
% \draw[
%     {Stealth[length=5pt]}-{Stealth[length=5pt]},
%     black!45,
%     line width=.45pt
% ]
%     (\xMinus+.06,-.5) -- (\xPlus-.06,-.5);

% \node[label,fill=white,inner sep=1.5pt]
%     at ({(\xMinus+\xPlus)/2},-.76)
%     {two intermediate lifts};

\end{tikzpicture}
\end{minipage}
\hspace{-20pt}
\begin{minipage}[t]{0.34\textwidth}
\centering
\begin{tikzpicture}[x=.9cm,y=.75cm]

    % Local coordinates on the raw correspondence
    \draw[black!25,line width=.4pt,-{Stealth[length=3pt]}]
        (0,-1.35) -- (0,1.45)
        node[above,font=\scriptsize] {$y$};

    \draw[black!25,line width=.4pt,-{Stealth[length=3pt]}]
        (-.08,0) -- (1.48,0)
        node[below,font=\scriptsize] {$s$};

    % Standard fold s=y^2
    \draw[
        black!65,
        line width=.7pt,
        domain=-1.15:1.15,
        samples=80,
        smooth,
        variable=\t
    ]
        plot ({\t*\t},{\t});

    \node[font=\scriptsize] at (1.20,1.6)
        {$\overline Z_{\alpha T}^{+}$};

    \node[font=\scriptsize] at (1.20,-1.6)
        {$\overline Z_{\alpha T}^{-}$};

    % Projection
    \draw[
        -{Stealth[length=3.3pt,width=4pt]},
        black!55,
        line width=.55pt
    ]
        (1.65,0) -- (3.15,0)
        node[midway,above=3pt,font=\scriptsize]
        {$\pi_{L\widetilde R}$};

    % Projected support
    \draw[black!65,line width=.7pt]
        (3.55,0) -- (5.85,0);

    \fill[black!65]
        (3.55,0) circle (1.5pt);

    \node[font=\scriptsize,above] at (4.75,0)
        {$Y_{\alpha T}^{\rm eff}$};

    \node[font=\scriptsize,below] at (3.55,0)
        {fold};

\end{tikzpicture}
\end{minipage}

\caption{ Left: for fixed external fields, the black curve schematically represents
 \eqref{eq:regular-loci-constraint}, while the horizontal line represents $\cos\alpha$. On the regular locus $|\cos\alpha|<R_{LR}$, their two intersections fix $x=\phi_L-\phi_I$ to two values. Right: local coalescence of the two regular branches. Their closures meet at $s=0$, where the projection has a single preimage.} 
\label{fig:Tdual-biconj-fold-and-solutions}
\end{figure}
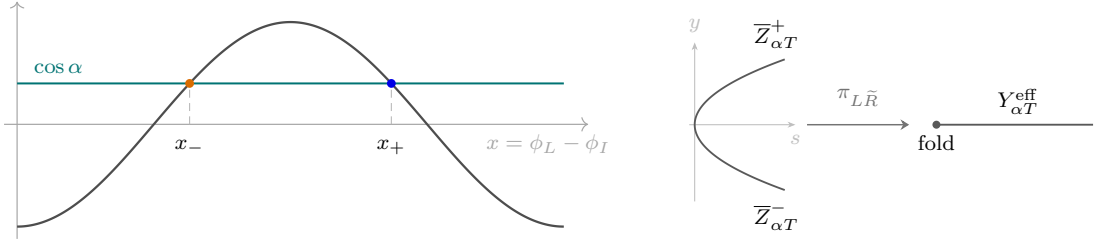

On the open locus
\begin{equation}\label{eq:Tdual-biconj-regular-locus}
U_{\alpha T} = \left\{ (g_L,\widetilde x_R)\in X_L\times\widetilde X_R \ \middle|\ |\cos\alpha|<R_{LR} \right\}\,, 
\end{equation}
the constraint has exactly two solutions for the intermediate fibre coordinate. For $R_{LR}>0$, write
\begin{equation}\label{eq:regular-loci-constraint}
A_{LR}\cos x+B_{LR}\sin x = R_{LR}\cos(x-\delta_{LR})\,, \quad \delta_{LR}:=\arg(A_{LR}+iB_{LR})\,,
\end{equation}
where $x=\phi_L-\phi_I$. The two solutions are
\begin{equation}\label{eq:Tdual-biconj-local-branches}
\phi_L-\phi_I = \delta_{LR} \pm \arccos\left(\frac{\cos\alpha}{R_{LR}}\right) \quad\bmod 2\pi\,.
\end{equation}
The problem defined by equation and its solutions are depicted in the left panel of Figure \ref{fig:Tdual-biconj-fold-and-solutions}.

The two sheets can be distinguished without choosing a branch of $\delta_{LR}$. Define
\begin{equation}
D_{LR}:=\frac{\partial}{\partial x}\left(A_{LR}\cos x+B_{LR}\sin x\right)=- A_{LR}\sin x+B_{LR}\cos x\,.
\end{equation}
Since
\begin{equation}
\left(A_{LR}\cos x+B_{LR}\sin x\right)^2+D_{LR}^2=A_{LR}^2+B_{LR}^2= R_{LR}^2\,,
\end{equation}
the constraint \eqref{eq:Tdual-biconj-explicit-constraint} implies
\begin{equation}
D_{LR}^2=R_{LR}^2-\cos^2\alpha\,.
\end{equation}
This is strictly positive on $U_{\alpha T}$. The loci $D_{LR}>0$ and $D_{LR}<0$ therefore define two smooth sheets $Z_{\alpha T}^{+}$ and $Z_{\alpha T}^{-}$ of the regular raw correspondence. This is in turn illustrated in the right panel of  Figure \ref{fig:Tdual-biconj-fold-and-solutions}, which details we will explain soon. For now, note that  equivalently,  $D_{LR}$ is the derivative of the constraint along the globally oriented intermediate Hopf-circle action. Its sign is therefore independent of the chosen local angular trivialisation. Their projections
\begin{equation}
\pi_\pm: Z_{\alpha T}^{\pm}\longrightarrow U_{\alpha T}
\end{equation}
are diffeomorphisms. Thus the regular correspondence consists of two globally distinguished branches; the multivaluedness of $\delta_{LR}$ does not exchange them. 

Equivalently, over every open set $V\subset U_{\alpha T}$,
\begin{equation}\label{eq:Tdual-biconj-local-cover}
\pi_{L\widetilde R}^{-1}(V) = V_+\sqcup V_-\,,
\end{equation}
where $\pi_\pm:V_\pm\to V$ are diffeomorphisms.

The two regular solutions and their coalescence at the boundary of $U_{\alpha T}$ are illustrated in the left panel of Figure \ref{fig:Tdual-biconj-fold-and-solutions}. For $0<|\cos\alpha|<1$, the locus $R_{LR}=|\cos\alpha|$ is a fold. Let $y$ denote the displacement of $x=\phi_L-\phi_I$ from the coalesced solution and set $s:=R_{LR}-|\cos\alpha|$. Expanding the constraint gives
\begin{equation}
s=\frac{|\cos\alpha|}{2}y^2+O(y^4)\,.
\end{equation}
Thus, locally, the projection has the standard form $s\simeq y^2$: there are two solutions for $s>0$, one coalesced solution for $s=0$ and no solution for $s<0$.

%%%%%%--------side-by-side tikz----------%%%%%%%
%%%%%%--------side-by-side tikz----------%%%%%%%

The regular part of the composition is summarised by
\begin{equation}\label{eq:Tdual-biconj-fusion-diagram}
\RawTdualBiconjFusionDiagram \;\;\FusionArrow\;\; \FusedTdualBiconjFusionDiagram\,.
\end{equation}
The diagram does not include the fold or the exceptional branch discussed below.

This differs from the Hopf defect fused with its reverse, considered in Section \ref{subsec:classical_fusion_Hopf}, where the generic factorisation fibre is a circle and its holonomy selects the $\mathbb Z_p$ branches. Here the generic fibre of $\pi_{L\widetilde R}$ consists of two isolated points. After separating them into the sheets $Z_{\alpha T}^{\pm}$, each restricted projection has point fibres, so no fibre-holonomy condition remains on the regular locus.

%%%%%%
\paragraph{Differential data on the regular branches.}

The raw intrinsic two-form on the fibre product is
\begin{equation}\label{eq:Tdual-biconj-raw-curvature}
F_{\alpha T}^{\rm raw} = \mathrm{pr}_{LI}^*F_\alpha + \mathrm{pr}_{I\widetilde R}^*F_T\,.
\end{equation}
Using \eqref{eq:Tdual-biconj-Falpha-twist} and \eqref{eq:Tdual-biconj-FT-twist}, the intermediate twists cancel:
\begin{equation}\label{eq:Tdual-biconj-raw-twist}
dF_{\alpha T}^{\rm raw} = p_L^*H-p_{\widetilde R}^*\widetilde H\,,
\end{equation}
where pullback to $Y_{\alpha T}^{\rm raw}$ is understood. Choose compatible local bulk potentials $db_L=H$, $db_I=H$ and $d\widetilde b_R=\widetilde H$. The local defect one-forms satisfy
\begin{equation}
F_\alpha = p_L^*b_L-p_I^*b_I+dA_\alpha\,, \qquad F_T = p_I^*b_I-p_{\widetilde R}^*\widetilde b_R+dA_T\,.
\end{equation}
Their sum gives
\begin{equation}\label{eq:Tdual-biconj-raw-one-form}
A_{\alpha T}^{\rm raw} = \mathrm{pr}_{LI}^*A_\alpha + \mathrm{pr}_{I\widetilde R}^*A_T \quad\text{with}\quad F_{\alpha T}^{\rm raw} = p_L^*b_L-p_{\widetilde R}^*\widetilde b_R + dA_{\alpha T}^{\rm raw}\,.
\end{equation}
%The support \eqref{} and intrinsic two-form \eqref{}  are defined classically for every $\alpha\in[0,\pi]$. 
%Statements involving the global prequantum connection require a prequantisable conjugacy class,
%\begin{equation}
%\alpha=\frac{\pi a}{k}\,, \qquad a\in\{0,\ldots,k\}\,.
%\end{equation}

Since $\pi_\pm$ are diffeomorphisms, the raw data transport directly to the two regular branches. They are uniquely determined by
\begin{equation}\label{eq:Tdual-biconj-branch-data}
\pi_\pm^*F_{\alpha T}^{\pm} = F_{\alpha T}^{\rm raw}\big|_{Z_{\alpha T}^{\pm}}\,, \qquad \pi_\pm^*A_{\alpha T}^{\pm} = A_{\alpha T}^{\rm raw}\big|_{Z_{\alpha T}^{\pm}}\,.
\end{equation}
Consequently,
\begin{equation}
F_{\alpha T}^{\pm} = p_L^*b_L-p_{\widetilde R}^*\widetilde b_R + dA_{\alpha T}^{\pm}\,.
\end{equation}
Thus the two solutions of the support constraint give two labelled classical branches on the same support $U_{\alpha T}$. They are  distinguished by their two-forms and connection data.

Each regular branch is a full-dimensional open subset of $X_L\times\widetilde X_R$. Hence
\begin{equation}
TU_{\alpha T} = T(X_L\times\widetilde X_R)\big|_{U_{\alpha T}}\,, \qquad (TU_{\alpha T})^\perp=0
\end{equation}
with respect to $G=g_L\oplus(-\widetilde g_R)$. The target-space topological defect conditions \cite{Kapustin:2010zc} reviewed in Section \ref{subsec:defect conf-topo-conditions} reduce to 
\begin{equation}\label{eq:Tdual-biconj-KS-conditions} 
\ker F_{\alpha T}^{\pm}=0\,, \qquad \left(G^{-1}F_{\alpha T}^{\pm}\right)^2=\mathbf 1\,.
\end{equation}
Since $\pi_\pm$ is a diffeomorphism, every external field variation on a branch lifts uniquely to the raw correspondence. The gluing conditions of the two constituent topological defects therefore compose without leaving an intermediate degree of freedom, and continuity of the stress tensor across the two interfaces implies continuity across the fused branch. By the equivalence reviewed in Section \ref{sec:review}, the transported data are topological and satisfy \eqref{eq:Tdual-biconj-KS-conditions}.

The regular fusion result is
\begin{equation}\label{eq:Tdual-biconj-regular-fusion}
% \left. \bigl( (B_\alpha,F_\alpha)\star(Y_T,F_T) \bigr) \right|_{\rm reg} = \bigsqcup_{\sigma=\pm} \bigl(U_{\alpha T}^{(\sigma)},F_{\alpha T}^{\sigma}\bigr)\,,\\
\left. (B_\alpha,F_\alpha) \star_{\mathrm{loc}} (Y_T,F_T) \right|_{\mathrm{reg}} = \bigsqcup_{\sigma=\pm} \bigl(U_{\alpha T}^{(\sigma)},F_{\alpha T}^{\sigma}\bigr)\,.
\end{equation}
where $U_{\alpha T}^{(\sigma)}$ denotes a labelled copy of the same open support $U_{\alpha T}$. When $(B_\alpha,F_\alpha)$ admits compatible global connection data, the same formula holds with $\star_{\mathrm{loc}}$ replaced by the full connection-level product $\star$, with the transported connections understood on both branches. The two copies labelled by $\pm$ carry the data transported from $Z_{\alpha T}^{\sigma}$. The closures of the two regular sheets meet at the fold, where $d\pi_{L\widetilde R}$ loses rank. The regular descent argument does not assign a smooth defect datum to this locus.

%%%%%%
\paragraph{Exceptional and degenerate loci.}

The only locus with a positive-dimensional factorisation fibre occurs when
\begin{equation}
\alpha=\frac{\pi}{2}\,, \quad R_{LR}=0\,.
\end{equation}
There the constraint \eqref{eq:Tdual-biconj-explicit-constraint} becomes independent of $\phi_I$, so the factorisation fibre is the full intermediate circle. The curvature and holonomy conditions must then be imposed along that circle.

As shown in Appendix \ref{sec:descent-T-dual-bibrane}, curvature descent restricts the exceptional locus to
\begin{equation}
k\phi_L+\widetilde\phi_R = \chi_0 \quad\bmod 2\pi\,. 
\end{equation}
If $C_{\pi/2}$ is prequantisable, then $k$ is even, and trivial fibre holonomy fixes $\chi_0=\frac{\pi k}{2} \;\bmod 2\pi$. The exceptional reduction therefore gives the topological graph defect
\begin{equation}
\bigl(Y_{\pi/2,T}^{\rm exc},0\bigr)\,,
\end{equation}
whose support is defined explicitly in \eqref{eq:Tdual-biconj-exceptional-branch}.

The special case bi-branes \eqref{eq:special-case-bibranes} provide two further checks. For $\alpha=0$, $C_0=\{\mathbf 1\}$ and $B_0$ is the identity defect. Hence
\begin{equation}
(B_0,F_0)\star(Y_T,F_T) \cong(Y_T,F_T)\,.
\end{equation}
For $\alpha=\pi$, one has $C_\pi=\{-\mathbf 1\}$, and hence $g_I=-g_L$. The projection of the raw correspondence is therefore the smooth support
\begin{equation}\label{eq:YpiT-support}
Y_{\pi T} := \left\{ (g_L,\widetilde x_R)\in X_L\times\widetilde X_R \ \middle|\ \pi_X(g_L)=\pi_{\widetilde X}(\widetilde x_R) \right\}\,.
\end{equation}
The restricted projection $\pi_{L\widetilde R}: Y_{\pi T}^{\rm raw}\longrightarrow Y_{\pi T}$ is a diffeomorphism, with inverse $s_\pi(g_L,\widetilde x_R)=(g_L,-g_L,\widetilde x_R)$.  The transported intrinsic two-form is consequently
\begin{equation}\label{eq:FpiT-transported}
F_{\pi T}:=s_\pi^*\left(\mathrm{pr}_{LI}^*F_\pi+\mathrm{pr}_{I\widetilde R}^*F_T\right)\,.
\end{equation}
The global connection data are transported by the same diffeomorphism. Thus
\begin{equation}\label{eq:Bpi-times-T}
(B_\pi,F_\pi)\star(Y_T,F_T)=(Y_{\pi T},F_{\pi T})\,,
\end{equation}
with the descended connection understood.

For $0<|\cos\alpha|<1$, regular points of $R_{LR}=|\cos\alpha|$ form a fold locus. Define $s:=R_{LR}-|\cos\alpha|$ and let $y$ measure the displacement of $x=\phi_L-\phi_I$ from the coalesced solution. Expanding the constraint gives
\begin{equation}
s=\frac{|\cos\alpha|}{2}y^2+O(y^4)\,.
\end{equation}
After a smooth rescaling of $s$, the projection therefore has the standard local form $s=y^2$. The right panel of Figure~\ref{fig:Tdual-biconj-fold-and-solutions} illustrates the local coalescence of the two branches. For $s>0$ there are two preimages, one on each sheet. At $s=0$ they coincide, while for $s<0$ there is no preimage.

%%%%%%--------table---------%%%%%%%
\begin{table}[t]
\centering
\small
\setlength{\tabcolsep}{3pt}
\renewcommand{\arraystretch}{1.5}

\begin{tabular}{
@{}
>{\centering\arraybackslash}m{0.18\linewidth}
>{\centering\arraybackslash}m{0.25\linewidth}
>{\centering\arraybackslash}m{0.24\linewidth}
>{\centering\arraybackslash}m{0.25\linewidth}
@{}
}
\multicolumn{4}{c}{\itshape WZW-compatible sector}\\[2pt]
\toprule
\cellcolor{black!3}$\star$
&
\cellcolor{black!3}$\begin{gathered}
(B_\beta,F_\beta)\\[-3pt]
0<\beta<\pi
\end{gathered}$
&
\cellcolor{black!3}$\bigl(Y_T,F_T\bigr)$
&
\cellcolor{black!3}$\bigl(Y_T,F_T\bigr)^\vee$
\\
\midrule

\cellcolor{black!3}
$\begin{gathered}
\bigl(B_\alpha,F_\alpha\bigr)\\[-3pt]
0<\alpha<\pi
\end{gathered}$
&
$\begin{gathered}
\bigl(B_\gamma,F_\gamma\bigr),\\[-2pt]
\gamma\in\mathcal A_{\alpha,\beta}^{\rm reg}
\end{gathered}$
&
$\displaystyle
\bigsqcup_{\sigma=\pm}
\bigl(
U_{\alpha T}^{(\sigma)},
F_{\alpha T}^{\sigma}
\bigr)$
&
---
\\

\cellcolor{black!3}$\bigl(B_0,F_0\bigr)$
&
$\bigl(B_\beta,F_\beta\bigr)$
&
$\bigl(Y_T,F_T\bigr)$
&
---
\\

\cellcolor{black!3}$\bigl(B_\pi,F_\pi\bigr)$
&
$\bigl(B_{\pi-\beta},F_{\pi-\beta}\bigr)$
&
$\bigl(Y_{\pi T},F_{\pi T}\bigr)$
&
---
\\

\cellcolor{black!3}$\bigl(Y_T,F_T\bigr)$
&
---
&
---
&
$\bigl(B_0,F_0\bigr)$
\\

\cellcolor{black!3}$\bigl(Y_T,F_T\bigr)^\vee$
&
$\displaystyle
\bigsqcup_{\sigma=\pm}
\bigl(
U_{\beta T}^{(\sigma)},
F_{\beta T}^{\sigma}
\bigr)^\vee$
&
$\bigl(\Delta_{L(k,1)},0\bigr)$
&
---
\\

\bottomrule
\end{tabular}

\vspace{1.4em}

\begin{tabular}{
@{}
>{\centering\arraybackslash}m{0.18\linewidth}
>{\centering\arraybackslash}m{0.25\linewidth}
>{\centering\arraybackslash}m{0.25\linewidth}
>{\centering\arraybackslash}m{0.25\linewidth}
@{}
}
\multicolumn{4}{c}{\itshape Fluxless Hopf sector}\\[2pt]
\toprule
\cellcolor{black!3}$\star$
&
\cellcolor{black!3}$\bigl(Y_{\delta_2},0\bigr)$
&
\cellcolor{black!3}$\bigl(Y_T,F_T\bigr)$
&
\cellcolor{black!3}$\bigl(Y_T^\vee,F_T^\vee\bigr)$
\\
\midrule

\cellcolor{black!3}$\bigl(Y_{\delta_1},0\bigr)$
&
$\bigl(Y_{\delta_1+\delta_2},0\bigr)$
&
$\begin{gathered}
\bigl(Y_T,F_T\bigr),\\[-2pt]
\mathrm{Hol}_{\rm flat}
=e^{-ip\delta_1}
\end{gathered}$
&
---
\\

\cellcolor{black!3}$\bigl(Y_T,F_T\bigr)$
&
---
&
---
&
$\displaystyle
\bigsqcup_{\eta\in\mathbb Z_p}
\bigl(Y_\eta,0\bigr)$
\\

\cellcolor{black!3}$\bigl(Y_T^\vee,F_T^\vee\bigr)$
&
$\begin{gathered}
\bigl(Y_T^\vee,F_T^\vee\bigr),\\[-2pt]
\mathrm{Hol}_{\rm flat}
=e^{-ip\delta_2}
\end{gathered}$
&
$\displaystyle
\bigsqcup_{\eta\in\mathbb Z_p}
\bigl(\widetilde Y_\eta,0\bigr)$
&
---
\\

\bottomrule
\end{tabular}

\vspace{1.4em}

\begin{tabular}{
@{}
>{\centering\arraybackslash}m{0.14\linewidth}
>{\centering\arraybackslash}m{0.50\linewidth}
>{\centering\arraybackslash}m{0.31\linewidth}
@{}
}
\multicolumn{3}{c}{\itshape TsT composition}\\[2pt]
\toprule
\cellcolor{black!3}Parameter
&
\cellcolor{black!3}Local reduction
&
\cellcolor{black!3}Global rank-one descent
\\
\midrule

\cellcolor{black!3}
$\zeta\in\mathbb R\setminus\{0\}$
&
$\begin{gathered}
\bigl(Y_T,F_T\bigr)
\star_{\mathrm{loc}}
\bigl(Y_{S_\zeta},0\bigr)
\star_{\mathrm{loc}}
\bigl(Y_{T_\zeta}^{\vee},F_{T_\zeta}^{\vee}\bigr)
{}=\bigl(Y_\zeta^{\mathrm{TsT}},F_\zeta\bigr)
\\[-2pt]
\qquad \text{pointwise topological}
\end{gathered}$
&
$\begin{gathered}
\displaystyle \zeta=\frac{2\pi}{N},\\[-2pt]
N\in\mathbb Z\setminus\{0\}\,,
\;\; N\mid p
\end{gathered}$
\\

\bottomrule
\end{tabular}

\caption{Summary of the classical composition laws derived in the
main text. Rows are fused with columns from left to right; dashes
denote incompatible intermediate backgrounds. Since several of the
defects separate the worldsheets of different sigma-model backgrounds,
the fusion operation in a given entry is defined only when the outgoing
target of the row defect agrees with the incoming target of the column
defect. The WZW tables display the regular branches. For the TsT
composition, global rank-one descent occurs only on the indicated
discrete locus.}
\label{tab:classical-defect-fusion}
\end{table}
%%%%%%--------table---------%%%%%%%

\raggedbottom
%%%%%%%%%%%%%%%%%%%%%%%%%%%%%%%%%%%%%%%%%%%%%%%%%%%%%%
%%%%%%%%%%%%%%         Conclusions        %%%%%%%%%%%%
%%%%%%%%%%%%%%%%%%%%%%%%%%%%%%%%%%%%%%%%%%%%%%%%%%%%%%
\section{Summary and conclusions}\label{sec:conclusions}

Classically, the fusion of sigma-model defects is usually approached by bringing two defect lines together, combining their worldsheet couplings and eliminating the intermediate fields. Although this simple approach might yield the resulting fusion products in individual examples, this action-level procedure does not provide an intrinsic rule for identifying the resulting target-space defect or resolving it into distinct branches.

Building on the target-space description of bi-branes \cite{Fuchs:2007fw} and the geometric characterisation of topological defects \cite{Kapustin:2010zc}, we formulated classical fusion as a target-space reduction. A defect is described by a submanifold of $X_L\times X_R$, an intrinsic two-form and connection data. The supports are composed by a fibre product over the intermediate target, while the differential data determine the loci on which the composition descends to the external targets. The resulting branches are then tested against the topologicality conditions of Section \ref{sec:review}. Our fusion prescription determines both the possible fusion branches and their local worldsheet one-form couplings, while retaining the global information that is lost if one considers only the projected support.

The results are collected in Table \ref{tab:classical-defect-fusion}. For the symmetry-preserving $SU(2)$ WZW bi-branes, the construction reproduces the known regular biconjugacy strata and their intrinsic two-forms \cite{Fuchs:2007fw}. For the Hopf-fibre T-duality interface, fusion with the reverse interface gives the finite family of fibre-shift defects, with the $\mathbb Z_p$ branches selected directly by the holonomy of the composite connection.

We then used our fusion prescription constructively. We obtained an explicit classical defect realisation of TsT, including its support, intrinsic two-form and local one-form action. The resulting correspondence is locally defined for a continuous deformation parameter, while full global connection-level fusion exists on the discrete locus determined in Section \ref{sec:TsT_fusion}. Finally, composing a WZW bi-brane with the Hopf T-duality interface gives the T-dual biconjugacy correspondences whose quantum counterparts were constructed in \cite{Sarkissian:2008dq}, together with their classical branch and connection data.\footnote{Since these defects connect different sigma-model backgrounds, a composition is defined only when the adjacent backgrounds agree. Further ordered products are collected in Appendix \ref{app:further-fusion-products}.}

These examples show, in particular, that non-invertible composition and its decomposition into several branches can already be visible classically. More generally, our construction turns classical fusion into a systematic geometric procedure that applies both to known fusion rules and to the construction of new defects, without requiring an ad hoc interpretation of the combined worldsheet action. Since the support, two-form and global connection data are retained throughout, the result also provides a concrete starting point for the subsequent quantisation of these defects.

\noindent
We conclude with a number of promising future directions:
\vspace{-5pt}
%%%%%%%%%
\paragraph{Singular branches and associativity.}
The fusion prescription we define here assumes that the fibre products and projections involved in fusion are sufficiently regular. The mixed WZW--T-duality fusion example already shows that fusion may also produce branches that meet or change dimension. Extending the construction to such singular loci should determine whether they contribute additional branches or multiplicities.
It also remains to establish associativity. Although the initial composition of the defect supports is associative up to a natural identification, the subsequent selection of admissible branches and descent of the connection data may actually depend on the order of fusion.

%%%%%%%%%
\paragraph{Quantisation and the role of twisted sectors.}
Ultimately, it would very interesting to utilise the geometric prescription to quantise to composition of defect operators. At the phase-space level this would require assigning Hilbert spaces to the sigma models and operators to the correspondences induced by the classical defects, compatibly with composition. The target-space data capturing the defect adopted  here provide part of the geometric input required for this quantisation, but, and crucially, do not by themselves identify a preferred or canonical quantisation scheme achieving this required compatible quantisation pipeline. Note though that for symmetry-preserving WZW bi-branes, the relation to Chern-Simons theory and quasi-Hamiltonian quantisation provides an established framework, in which fusion is mapped to the Verlinde product \cite{Fuchs:2007fw,Axelrod:1989xt,meinrenken2011quantization}.

The Hopf T-duality defects on the other hand present a different problem. Their quantum description is related to an orbifold construction and therefore involves twisted sectors \cite{Fuchs:2007tx,Dijkgraaf:1989hb}. Reproducing this description through the Chern-Simons route may require enlarging the relevant moduli space to include flat connections with prescribed discrete monodromy, or an appropriate orbifold moduli space. A concrete test would be to show that the quantisation of the classical $\mathbb Z_p$ branches found here gives the corresponding decomposition of the product of the T-duality operator with its reverse.  

%%%%%%%%%
\paragraph{Higher groups, Abelian dualities and coset models.}
Although the examples considered here are based on $SU(2)$, the geometric prescription is not restricted to this group. Its extension to higher-rank compact groups would provide a direct test of the descent conditions for higher-dimensional factorisation fibres and more involved singular strata. Similarly, replacing the Hopf circle by higher-dimensional torus actions should lead to a geometric treatment of more general Abelian duality defects and their compositions. For gauged WZW models and coset targets, the construction must be made compatible with gauge reduction, requiring an equivariant version of both the support and connection-level descent. Such an extension would also be a first step towards applying the formalism to group and coset sigma models of holographic interest. Non-compact groups, supercosets and backgrounds with Ramond-Ramond flux would require further ingredients beyond the bosonic NS-flux setting treated here.

%%%%%%%%%%%%%%%%%%%%%%%%%%%%%%%%%%%%%%%%%%%%%%%%%%%%%%
\subsubsection*{Acknowledgments}
We would like to thank Alex S. Arvanitakis, Jordi Gaset Rif\`a, Christian Northe and Peter Schupp for interesting discussions. We also thank the participants of the 6th Mini symposium on Physics and Geometry, where preliminary results of this work were presented, for useful discussions. We further thank the participants of the Global Categorical Symmetries 2026 conference week for helpful exchanges. C.W. is supported by the program PIPF-2024/TEC-34293 from Comunidad de Madrid.

\appendix
\addtocontents{toc}{\protect\setcounter{tocdepth}{1}}
%%%%%%%%%%%%%%%%%%%%%%%%%%%%%%%%%%%%%%%%%%%%%%%%
%%%%%%    Appendix top cond Hopf defect   %%%%%%
%%%%%%%%%%%%%%%%%%%%%%%%%%%%%%%%%%%%%%%%%%%%%%%%
\section{Additional details topologicality conditions}

%%%%%%%%%%%%%%%%%%%%%%%%%%%%%%%%%%%%%%%%%%%%%%%%
%%%%%%%%%%%%%%%%%%%%%%%%%%%%%%%%%%%%%%%%%%%%%%%%
\subsection{Topologicality of the Hopf defect} \label{app:Hopf_topologicality_details}

In this appendix, we verify topological conditions reviewed in Section \ref{subsec:defect conf-topo-conditions} for the Hopf-fibre defect in Section \ref{subsec:Hopf_T_duality}. On the correspondence space $Y_T$, we introduce the one-forms 
\begin{equation}
\xi_L:=2d\phi-\cos\theta\,d\psi\,, \quad \Omega_{S^2}:=\sin\theta\,d\theta\wedge d\psi =d\xi_L\,.
\end{equation}
The pullback of the neutral metric $G=g_L\oplus(-g_R)$ to $Y_T$ is
\begin{equation}
\left.G\right|_{TY_T} = \frac{R_L^2}{8}\,\xi_L^2 -\frac{R_R^2}{2}\,d\widetilde\phi^2 \,.
\end{equation}
The base contributions cancel because the two projections from $Y_T$ have the same $(\theta,\psi)$ coordinates. Then, the most general intrinsic two-form compatible with the symmetries is
\begin{equation}
F_T = a\,\xi_L\wedge d\widetilde\phi +b\,\Omega_{S^2}\,,
\end{equation}
where $a$ and $b$ are constants. Since
\begin{equation}
H_L=0\,, \qquad H_R= \frac{p}{4\pi}\, \Omega_{S^2}\wedge d\widetilde\phi\,,
\end{equation}
the curvature condition $dF_T=p_L^*H_L-p_R^*H_R$ fixes $a=-\frac{p}{4\pi}$ while leaving $b$ undetermined.

Define the two-dimensional distribution
\begin{equation}
\mathcal K= \ker\xi_L\cap\ker d\widetilde\phi = \mathrm{span} \left\{ \partial_\theta,\, \partial_\psi+\frac12\cos\theta\,\partial_\phi \right\}\,.
\end{equation}
Restricting the neutral metric shows that $\mathcal K\subseteq(TY_T)^\perp$. Since $Y_T$ has codimension two in $X_L\times X_R$, both distributions have rank two, and therefore one obtains
\begin{equation}
\mathcal K=(TY_T)^\perp\,.
\end{equation}
Hence one can conclude that $Y_T$ is coisotropic. The kernel condition $(TY_T)^\perp\subseteq\ker F_T$ forces $b=0$, since the restriction of $\Omega_{S^2}$ to $\mathcal K$ is nondegenerate. Consequently,
\begin{equation}\label{eq:Hopf_topological_F}
F_T =-\frac{p}{4\pi}\,\xi_L\wedge d\widetilde\phi=-\frac{p}{2\pi}\left(d\phi-\frac12\cos\theta\,d\psi\right)\wedge d\widetilde\phi\,,
\end{equation}
and $\ker F_T=(TY_T)^\perp$.  It is left to impose the quotient condition. On $\mathcal Q_{Y_T} = TY_T/(TY_T)^\perp$ use the coframe $(\xi_L,d\widetilde\phi)$. The descended tensors are represented by
\begin{equation}
\widetilde G = \begin{pmatrix}
R_L^2/8 & 0\\
0 & -R_R^2/2
\end{pmatrix}\,, \qquad \widetilde F = \begin{pmatrix}
0 & -p/(4\pi)\\
p/(4\pi) & 0
\end{pmatrix}\,. 
\end{equation}
A direct computation gives
\begin{equation}
\left(\widetilde G^{-1}\widetilde F\right)^2 = \frac{p^2}{\pi^2R_L^2R_R^2}\, \mathrm{id}_{\mathcal Q_{Y_T}}\,.
\end{equation}
The topological defect condition reviewed in Section \ref{subsec:defect conf-topo-conditions} therefore requires
\begin{equation}\label{eq:Hopf_parameter_radius_condition}
R_LR_R=\frac{p}{\pi}\,.
\end{equation}
With the metrics used above, the physical radii of the two fibre coordinates are $r_L=\frac{R_L}{\sqrt2}$ and $r_R=\frac{R_R}{\sqrt2}$. Equation \eqref{eq:Hopf_parameter_radius_condition} is therefore equivalently
\begin{equation}
r_Lr_R=\frac{p}{2\pi}\,,
\end{equation}
which is the usual T-duality condition in the conventions of the main text.

%%%%%%%%%%%%%%%%%%%%%%%%%%%%%%%%%%%%%%%%%%%%%%%%
%%%%%%%%%%%%%%%%%%%%%%%%%%%%%%%%%%%%%%%%%%%%%%%%
\subsection{Exceptional descent in the mixed WZW--T-duality fusion}\label{sec:descent-T-dual-bibrane}

This appendix completes the descent analysis of the mixed composition studied in Section~\ref{subsec:Tdual-biconjugacy-defects}. Its raw support, projection and intrinsic two-form are defined in \eqref{eq:Tdual-biconj-raw-support}, \eqref{eq:Tdual-biconj-projection} and \eqref{eq:Tdual-biconj-raw-curvature}. On the regular locus \eqref{eq:Tdual-biconj-regular-locus}, the projection has discrete fibres and the data are transported branchwise. Here we treat the exceptional circle fibre occurring at
\begin{equation}
\alpha=\frac{\pi}{2}\,, \quad R_{LR}=0\,.
\end{equation}
At this locus the biconjugacy constraint becomes independent of $\phi_I$, rather than developing the fold shown in Figure \ref{fig:Tdual-biconj-fold-and-solutions}. Locally, the external base points are antipodal,
\begin{equation}\label{eq:Tdual-biconj-antipodal-base}
\theta_L=\pi-\theta_R\,, \qquad \psi_L=\psi_R+\pi \;\;\bmod 2\pi\,,
\end{equation}
and the factorisation fibre is the intermediate Hopf circle generated by $V=\partial_{\phi_I}$.

For the Hopf T-duality defect $(Y_T,F_T)$ used in the mixed composition, introduce the bundle connection one-forms
\begin{equation}
\xi_I=d\phi_I-\frac12\cos\theta_R\,d\psi_R\,,\qquad \widetilde\xi_R=d\widetilde\phi_R-\frac{k}{2}\cos\theta_R\,d\psi_R\,,
\end{equation}
which satisfy $d\widetilde\xi_R=k\,d\xi_I$. The intrinsic two-form of the Hopf defect is locally
\begin{equation}\label{eq:WZW-Hopf-form-exceptional-descent}
F_T=-\frac{1}{2\pi}\xi_I\wedge\widetilde\xi_R\,.
\end{equation}
Since $\omega_{\pi/2}=0$, the WZW bi-brane form on $B_{\pi/2}$ is
\begin{equation}
F_{\pi/2}=-\frac{k}{2}\left\langle q^{-1}dq\wedge dg_I\,g_I^{-1}\right\rangle\,,\quad g_L=qg_I\,.
\end{equation}
Let $t=i\sigma_3$, so that $\partial_{\phi_I}g_I=g_It$. Invariance of the bilinear form gives
\begin{equation}
\iota_{\partial_{\phi_I}}F_{\pi/2}=\frac{k}{2}\left\langle t,g_L^{-1}dg_L\right\rangle=-\frac{k}{2\pi}\xi_L\,,
\end{equation}
where $\xi_L=d\phi_L-\frac12\cos\theta_L\,d\psi_L$. Moreover we have that  $\iota_{\partial_{\phi_I}}F_T=-\frac{1}{2\pi}\widetilde\xi_R$ and hence
\begin{equation}
\iota_{\partial_{\phi_I}}F_{\pi/2,T}^{\rm raw}=-\frac{1}{2\pi}\left(k\xi_L+\widetilde\xi_R\right)\,.
\end{equation}
On the antipodal-base locus \eqref{eq:Tdual-biconj-antipodal-base},
\begin{equation}
k\xi_L+\widetilde\xi_R= d\left(k\phi_L+\widetilde\phi_R\right)\,.
\end{equation}
The curvature-descent condition
\eqref{eq:fusion-cond-horiz} therefore restricts the exceptional
locus to the level sets
\begin{equation}\label{eq:Tdual-biconj-exceptional-level}
k\phi_L+\widetilde\phi_R
=
\chi_0
\quad\bmod 2\pi.
\end{equation}
Curvature descent leaves $\chi_0$ undetermined; it is fixed by the
fibre-holonomy condition \eqref{eq:fibre-holonomy}.

Assume that $C_{\pi/2}$ carries the standard level-$k$ prequantum
datum. Writing $\alpha=\frac{\pi a}{k}$, the condition $\alpha=\pi/2$ requires $a=k/2$, and therefore even $k$. Under the map $q=g_Lg_I^{-1}$, the intermediate fibre circle traces an equator of $C_{\pi/2}$. The conjugacy-class datum contributes the holonomy $\exp(i\pi a)=(-1)^a$. Evaluating the remaining part of the composite connection gives the raw fibre holonomy
\begin{equation}\label{eq:Tdual-biconj-exceptional-holonomy}
\mathrm{Hol}_{\phi_I}=(-1)^a\exp\left[-i\left(k\phi_L+\widetilde\phi_R\right)\right]\,.
\end{equation}
Here the dependence on $\widetilde\phi_R$ is the Poincar\'e contribution. Reversing the fibre orientation replaces \eqref{eq:Tdual-biconj-exceptional-holonomy} by its inverse and leaves the trivial-holonomy condition unchanged. On the level set \eqref{eq:Tdual-biconj-exceptional-level}, trivial holonomy requires
\begin{equation}
\chi_0=\pi a=\frac{\pi k}{2}\quad\bmod 2\pi\,.
\end{equation}
Equivalently, $\chi_0=0$ for $k=0\bmod4$ and $\chi_0=\pi$ for $k=2\bmod4$.

The exceptional reduction therefore produces the three-dimensional support
\begin{equation}\label{eq:Tdual-biconj-exceptional-branch}
Y_{\pi/2,T}^{\rm exc}= \left\{
\begin{aligned}
\theta_R&=\pi-\theta_L,\\
\psi_R&=\psi_L-\pi,\\
\widetilde\phi_R&=\pi a-k\phi_L
\end{aligned}\quad\bmod 2\pi\right\}\subset X_L\times\widetilde X_R\,.
\end{equation}
Restricting the raw two-form to the corresponding exceptional admissible locus gives zero. It therefore descends to
\begin{equation}
\bigl(Y_{\pi/2,T}^{\rm exc},0\bigr)\,.
\end{equation}
Note that we can geometrically identify $Y_{\pi/2,T}^{\rm exc}$ with the graph of the (degree-$k$ bundle) map from $SU(2)$ to $L(k,1)$ covering the antipodal map of $S^2$. This map is a local isometry for the T-dual metrics, so the resulting graph defect is topological.

%%%%%%%%%%%%%%%%%%%%%%%%%%%%%%%%%%%%%%%%%%%%%%%%
%%%%%%%%% Appendix remaining fusions %%%%%%%%%%%
%%%%%%%%%%%%%%%%%%%%%%%%%%%%%%%%%%%%%%%%%%%%%%%%
\section{Further fusion products}\label{app:further-fusion-products}

In this appendix we derive the fusion products that follow directly from the calculations in the main text. We retain the source and target of every defect, since products with mismatched intermediate sigma models are not defined.

For a defect $(Y,F)\subset X_1\times X_2$, its orientation reverse is
\begin{equation}\label{eq:orientation-reversed-defect}
(Y,F)^\vee = \bigl(\tau(Y),-\tau^*F\bigr) \subset X_2\times X_1\,,
\end{equation}
where $\tau(x_1,x_2)=(x_2,x_1)$. For local connection one-forms, $A^\vee=-\tau^*A$. Orientation reversal reverses the order of fusion:
\begin{equation}\label{eq:reversal-reverses-fusion}
\bigl((Y_{12},F_{12})\star(Y_{23},F_{23})\bigr)^\vee = (Y_{23},F_{23})^\vee\star(Y_{12},F_{12})^\vee\,.
\end{equation}

%%%%%%%%%%%%%%%%%%%%%%%%%%%%%%%%%%%%%%%%%%%%%%%%
%%%%%%%%%%%%%%%%%%%%%%%%%%%%%%%%%%%%%%%%%%%%%%%%
\subsection{Fibre shifts and the Hopf}

Let $(Y_\delta,0)$ be the fibre-shift defect
\begin{equation}
Y_\delta = \left\{ \theta_L=\theta_I,\; \psi_L=\psi_I,\; \phi_I=\phi_L+\delta \right\} \subset X_L\times X_I\,, \quad \delta\in\mathbb R/2\pi\mathbb Z\,.
\end{equation}
We compose it with the degree-$p$ Hopf defect $(Y_T,F_T)\subset X_I\times\widetilde X_R$. In the local gauge used in the main text,
\begin{equation}\label{eq:app-Hopf-local-data}
F_T = -\frac{p}{2\pi} \left( d\phi_I-\frac{1}{2}\cos\theta\,d\psi \right) \wedge d\widetilde\phi_R\,, \quad A_T= -\frac{p}{2\pi}\phi_I\,d\widetilde\phi_R\,.
\end{equation}
Eliminating $\phi_I$ gives the original Hopf support, while $d\phi_I=d\phi_L$. Consequently, the intrinsic two-form remains $F_T$, but the local connection becomes
\begin{equation}\label{eq:shift-times-T-connection}
A_{\delta T} = -\frac{p}{2\pi} (\phi_L+\delta)\,d\widetilde\phi_R = A_T(\phi_L,\tilde{\phi}_R)-\frac{p\delta}{2\pi}\,d\widetilde\phi_R\,.
\end{equation}
Thus fusion with a generic fibre shift preserves the support and intrinsic two-form of the Hopf defect, but modifies its global connection data. Let $L_\delta\to Y_T$ denote the flat line bundle whose holonomy around the dual circle is
\begin{equation}\label{eq:shift-times-T-holonomy}
\mathrm{Hol}_{\widetilde S^1}(L_\delta) =\exp(-ip\delta)\,.
\end{equation}
With the trivial flat connection chosen on the shift defect, the full connection-level fusion is
\begin{equation}\label{eq:generic-shift-times-T}
(Y_\delta,0,\nabla_\delta) \star (Y_T,F_T,\nabla_T) = \bigl(Y_T,F_T,\nabla_T\otimes L_\delta\bigr)\,.
\end{equation}
Since the Hopf correspondence has no additional independent one-cycles, $L_\delta$ is trivial precisely when $p\delta\in2\pi\mathbb Z$. Consequently, for the discrete shifts $\delta=2\pi\eta/p$ one obtains
\begin{equation}\label{eq:discrete-shift-absorbed-by-T}
\left( Y_{\frac{2\pi}{p}\eta}, 0, \nabla_{\frac{2\pi}{p}\eta} \right)\star(Y_T,F_T,\nabla_T)=(Y_T,F_T,\nabla_T)\,,\qquad\eta\in\mathbb Z_p\,.
\end{equation}
For a generic $\delta$, the fused defect has the same support and intrinsic two-form as $(Y_T,F_T)$, but is not isomorphic to it as a defect with connection data because $L_\delta$ has non-trivial holonomy.

The corresponding result for the reversed  follows in the same way. Since $A_T^\vee = \frac{p}{2\pi}\phi_I\,d\widetilde\phi_L$ and the second defect imposes $\phi_R=\phi_I+\delta$, one obtains
\begin{equation}
A_{T^\vee\delta} = \frac{p}{2\pi}(\phi_R-\delta)\,d\widetilde\phi_L = A_T^\vee-\frac{p\delta}{2\pi}\,d\widetilde\phi_L\,.
\end{equation}
Hence
\begin{equation}
(Y_T,F_T)^\vee\star (Y_{\frac{2\pi}{p}\eta},0) =(Y_T,F_T)^\vee\,, \quad \eta\in\mathbb Z_p\,,
\end{equation}
whereas a generic shift again changes the flat connection by a holonomy $\exp(-ip\delta)$.

%%%%%%%%%%%%%%%%%%%%%%%%%%%%%%%%%%%%%%%%%%%%%%%%
%%%%%%%%%%%%%%%%%%%%%%%%%%%%%%%%%%%%%%%%%%%%%%%%
\subsection{Fusion in the opposite order}

We next compute the composition
\begin{equation}
(Y_T,F_T)^\vee\star(Y_T,F_T)\,,
\end{equation}
which is an endodefect of the T-dual background. Its raw support is
\begin{equation}
Y_{\vee T}^{\mathrm{raw}} = Y_T^\vee\times_XY_T \subset \widetilde X_L\times X_I\times\widetilde X_R\,.
\end{equation}
Writing $\xi_I = d\phi_I-\frac{1}{2}\cos\theta\,d\psi$ and $\Delta\widetilde\phi=\widetilde\phi_L-\widetilde\phi_R$, the raw two-form is
\begin{equation}\label{eq:reverse-Hopf-square-raw-form}
F_{\vee T}^{\mathrm{raw}}=\frac{p}{2\pi}\,\xi_I\wedge d\Delta\widetilde\phi\,.
\end{equation}
The vertical direction of the projection to the external targets is generated by $\partial_{\phi_I}$, and therefore
\begin{equation}
\iota_{\partial_{\phi_I}}F_{\vee T}^{\mathrm{raw}} = \frac{p}{2\pi}\,d\Delta\widetilde\phi\,.
\end{equation}
horizontality of the raw intrinsic two-form restricts the raw correspondence to the level sets $\Delta\widetilde\phi=\text{constant}$.  The raw local connection may be written as $A_{\vee T}^{\mathrm{raw}} = \frac{p}{2\pi}\phi_I\,d\Delta\widetilde\phi$. Locally,
\begin{equation}\label{eq:Araw_rev}
A_{\vee T}^{\mathrm{raw}} = -\frac{p}{2\pi}\Delta\widetilde\phi\,d\phi_I + d\left( \frac{p}{2\pi}\phi_I\Delta\widetilde\phi \right)\,.
\end{equation}
We thus have that for $A_{\rm raw}=p\phi_Id \Delta \tilde \phi/(2\pi)$,  under the change $\phi_I\mapsto \phi_I+2\pi$ is shifted by $pd\Delta\tilde \phi$. Hence the associated holonomy is ${\rm Hol}=e^{i\oint A}=e^{-ip\Delta \tilde \phi}$. Trivial fibre holonomy gives $\Delta\widetilde\phi = -\frac{2\pi}{p}\eta$, for $\eta\in\mathbb Z_p$.  Define then the dual fibre-shift branches by
\begin{equation}\label{eq:dual-shift-branches}
\widetilde Y_\eta = \left\{ \theta_L=\theta_R,\quad \psi_L=\psi_R,\quad \widetilde\phi_L = \widetilde\phi_R-\frac{2\pi}{p}\eta \right\} \subset \widetilde X_L\times\widetilde X_R\,.
\end{equation}
The restriction of \eqref{eq:reverse-Hopf-square-raw-form} to each branch vanishes, so the descended intrinsic two-form is zero. Thus
\begin{equation}
(Y_T,F_T)^\vee\star(Y_T,F_T) = \bigsqcup_{\eta\in\mathbb Z_p} (\widetilde Y_\eta,0)\,.
\end{equation}
In the WZW-compatible sector, the pairing in \eqref{eq:WZW-Hopf-form-exceptional-descent} has coefficient $1/(2\pi)$, while $k$ is the Chern number of the dual Hopf bundle. With inverse connection data, the intermediate-circle holonomies therefore picks only the zero relative shift. Consequently,
\begin{equation}
    (Y_T,F_T)\star(Y_T,F_T)^\vee=(B_0,F_0)\,, \quad (Y_T,F_T)^\vee\star(Y_T,F_T) =\bigl(\Delta_{L(k,1)},0\bigr)\,,
\end{equation}
where $\Delta_{L(k,1)}$ denotes the diagonal in $L(k,1)\times L(k,1)$.

%%%%%%%%%%%%%%%%%%%%%%%%%%%%%%%%%%%%%%%%%%%%%%%%
\subsection{The reversed mixed fusion}

The remaining mixed product in the WZW-compatible sector is
\begin{equation}
(Y_T,F_T)^\vee\star(B_\beta,F_\beta)\,.
\end{equation}
Its raw correspondence is
\begin{equation}\label{eq:reversed-mixed-raw-support}
Y_{T^\vee\beta}^{\mathrm{raw}} = Y_T^\vee\times_{X_I}B_\beta
\subset \widetilde X_L\times X_I\times X_R,
\end{equation}
with $F_{T^\vee\beta}^{\mathrm{raw}} = \mathrm{pr}_{\widetilde L I}^*F_T^\vee + \mathrm{pr}_{IR}^*F_\beta$. For $SU(2)$, $C_\beta^{-1}=C_\beta$, and hence the symmetry-preserving bi-brane is self-reversed:
\begin{equation}
(B_\beta,F_\beta)^\vee=(B_\beta,F_\beta)\,.
\end{equation}
Equation \eqref{eq:reversal-reverses-fusion} therefore gives
\begin{equation}\label{eq:reversed-mixed-fusion}
% (Y_T,F_T)^\vee\star(B_\beta,F_\beta) \simeq \bigl( (B_\beta,F_\beta)\star(Y_T,F_T) \bigr)^\vee\,.\\
\left. (Y_T,F_T)^\vee \star_{\mathrm{loc}} (B_\beta,F_\beta) \right|_{\mathrm{reg}}
= \left( \left. (B_\beta,F_\beta) \star_{\mathrm{loc}} (Y_T,F_T) \right|_{\mathrm{reg}} \right)^\vee\,.
\end{equation}
When the input data admit compatible global connections, the same identity holds with $\star_{\mathrm{loc}}$ replaced by $\star$.

Thus the reversed mixed fusion has the same regularity properties as the mixed fusion studied in the main text. The regular locus separates into the two globally distinguished components $D_{LR}>0$ and $D_{LR}<0$.

%%%%%%%%%
\paragraph{Resulting closure.}

Hence we complete the remaining multiplications of table \ref{tab:classical-defect-fusion}
\begin{align}
\left. (Y_T,F_T)^\vee \star_{\mathrm{loc}} (B_\beta,F_\beta) \right|_{\mathrm{reg}}
&= \left( \left. (B_\beta,F_\beta) \star_{\mathrm{loc}} (Y_T,F_T) \right|_{\mathrm{reg}} \right)^\vee\,.\\
(Y_T,F_T)\star(Y_T,F_T)^\vee &= \bigsqcup_{\eta\in\mathbb Z_p}(Y_\eta,0)\,, \\
(Y_T,F_T)^\vee\star(Y_T,F_T) &= \bigsqcup_{\eta\in\mathbb Z_p}(\widetilde Y_\eta,0)\,.
\end{align}
For the fluxless Hopf sector, the additional products are
\begin{align}
(Y_\delta,0,\nabla_\delta) \star (Y_T,F_T,\nabla_T) &= (Y_T,F_T,\nabla_T\otimes L_\delta)\,,\\
(Y_T^\vee,F_T^\vee,\nabla_T^\vee) \star (Y_\delta,0,\nabla_\delta) &= (Y_T^\vee,F_T^\vee,  \nabla_T^\vee\otimes\widetilde L_\delta)\,,
\end{align}
Here $L_\delta$ and $\widetilde L_\delta$ denote the corresponding flat line data, whose holonomy around the dual circle is $e^{-ip\delta}$. They are gauge-trivial precisely when $\delta=2\pi\eta/p$. For $\delta=2\pi\eta/p$, the additional flat connection is gauge-trivial and these reduce to
\begin{align}
(Y_{\frac{2\pi}{p}\eta},0)\star(Y_T,F_T) &=(Y_T,F_T)\,, \\ 
(Y_T,F_T)^\vee\star (Y_{\frac{2\pi}{p}\eta},0) &=(Y_T,F_T)^\vee\,.
\end{align}

%%%%%%%%%%%%%%%%%%%%%%%%%%%%%%%%%%%%%%%%%%%%%%%%%%%%%%
%%%%%%%%%%%%%%         Bibliography       %%%%%%%%%%%%
%%%%%%%%%%%%%%%%%%%%%%%%%%%%%%%%%%%%%%%%%%%%%%%%%%%%%%
\bibliographystyle{JHEP}
\bibliography{biblio}   
\end{document}